\documentclass[journal]{IEEEtran}
\usepackage{amsmath,amsfonts}
\usepackage{algorithmic}
\usepackage{algorithm}
\usepackage{array}
\usepackage{textcomp}
\usepackage{stfloats}
\usepackage{url}
\usepackage{verbatim}
\usepackage{graphicx}
\usepackage{cite}
\usepackage{comment}
\usepackage{makecell}
\usepackage{tabularx}
\usepackage{amsmath}
\usepackage{subfigure}
\usepackage[subfigure]{tocloft}
\usepackage{comment}
\usepackage{multirow}                 
\usepackage{multicol}                 
\usepackage{multirow}                
\usepackage{makecell}                 
\usepackage{booktabs}                 
\usepackage[table,xcdraw]{xcolor} 
\usepackage{booktabs} 

\usepackage{amsmath}

\usepackage{nomencl}
\usepackage{xstring}
\makenomenclature

\usepackage{etoolbox}
\renewcommand\nomgroup[1]{%
  \item[\bfseries
  \ifstrequal{#1}{A}{Sets}{%
  \ifstrequal{#1}{B}{Parameters}{%
  \ifstrequal{#1}{C}{Variables}{%
  \ifstrequal{#1}{D}{Abbreviations}{}}}}%
]}

\begin{document}

\title{
Performance-Aware Control of Modular Batteries For Fast Frequency Response
}

\author{
\IEEEauthorblockN{
Yutong He, 
Guangchun Ruan, \textit{Member, IEEE},
Haiwang Zhong, \textit{Senior Member, IEEE}
}

\thanks{
This work was supported by Key-Area Research and Development Program of Guangdong Province under Grant No. 2023B0909040003. \textit{(Corresponding author: Haiwang Zhong.)}

Yutong He and Haiwang Zhong are with the State Key Laboratory of Power System Operation and
Control, Department of Electrical Engineering, Tsinghua University, Beijing
100084, China (e-mail: hyt22@mails.tsinghua.edu.cn; zhonghw@tsinghua.edu.cn).

Guangchun Ruan is with the Lab for Information and Decision Systems,
Massachusetts Institute of Technology, Cambridge, MA 02139 USA (e-mail:
gruan@mit.edu).



}
\thanks
}



\maketitle

\begin{abstract}
Modular batteries can be aggregated to deliver frequency regulation services for power grids. 
Although utilizing the idle capacity of battery modules is financially attractive, it remains challenging to consider the heterogeneous module-level characteristics such as dynamic operational efficiencies and battery degradation.
In addition, real-time decision making within seconds is required to enable fast frequency response.
In order to address these issues, this paper proposes a performance-aware scheduling approach for battery modules to deliver fast frequency response~(FFR) support.
In particular, the conduction loss and switching loss of battery packs as well as converters are captured within a mix-integer quadratic constrained program~(MIQCP).
The cycle-based aging model identifies the aging cost of battery modules during frequent cycling by introducing the aging subgradient calculation and linearization.
Case studies based on real-world battery data show that the proposed scheduling approach can effectively reduce power loss cost by nearly 28\%-57\% and battery aging cost by 4\%-15\% compared to conventional methods, which can also enhance the SoC balance. 


\end{abstract}

\begin{IEEEkeywords}
battery control,
energy storage systems,
distributed energy resource,
frequency support,
electricity market
\end{IEEEkeywords}

\nomenclature[A, 01]{\(\mathcal{N}\)}{Set of modules in one system}
\nomenclature[A, 02]{\(\mathcal{K}\)}{Set of battery systems}
\nomenclature[A, 03]{\(\mathcal{T}\)}{Set of scheduling time intervals}
\nomenclature[A, 04]{$\eta _{}^{{\rm{ch}}}$}{Set of modular charging efficiency}
\nomenclature[A, 04]{$\eta _{}^{{\rm{dch}}}$}{Set of modular discharging efficiency}
\nomenclature[A, 05]{$SOC$}{Set of modular SoC}
\nomenclature[B, 01]{$C_i^{{\rm{Coss}}}$}{The output capacitance of converter}
\nomenclature[B, 01]{$C_{{\rm{bess}}}^{{\rm{bid}}}$}{The bidding capacity of aggregated battery system}

\nomenclature[B, 02]{$DC{R_{k,i}}$}{The equivalent resistance of the inductance in the converter} 
\nomenclature[B, 02]{${E_{k,i}}$}{The maximum energy capacity of battery
module \(i\) in system \(k\).}

\nomenclature[B, 03]{$f_{sw}$}{The switching frequency of MOSFET}
\nomenclature[B, 04]{$N_{100\% }^{cycle}$}{The number of available battery cycles at 100\% cycle depth}

\nomenclature[B, 04]{$P_{\rm{t}}^{{\rm{RegD}}}$}{The value of regulation command during time span $t$}
\nomenclature[B, 04]{${Q_{{\rm{rr}},k,i}}$}{The diode reverse recovery charge}
\nomenclature[B, 04]{$Q_{{\rm{g1}},k,i}$,$Q_{{\rm{g2}},k,i}$}{The total gate
charge of the two MOSFETs in the converter.}

\nomenclature[B, 05]{$R_{{\rm{bat,}}k,i}^{}$}{The internal resistance of the battery pack}
\nomenclature[B, 06]{$R_{{\rm{on,}}k,i}^{}$}{The on-resistance of MOSFET}
\nomenclature[B, 06]{${r_t}$}{The value of regulation signals during time span $t$}

\nomenclature[B, 06]{$SoC_{k,i,\min }$,$SoC_{k,i,\max }$}{The minimum/maximum SoC}

\nomenclature[B, 07]{${{t_r}}$, ${{t_f}}$}{The rising/falling time of the switching transition}
\nomenclature[B, 08]{${V_{{\rm{dc}},k,i}}$}{The DC bus side voltage of module \(i\)}
\nomenclature[B, 09]{${V_{{\rm{DS}},k,i}}$}{The drain-source voltage of module \(i\)}
\nomenclature[B, 10]{${V_{{\rm{GS}},k,i}}$}{The gate-source voltage of module \(i\)}
\nomenclature[B, 11]{$\pi _{k,i}^{{\rm{bat\_cost}}}$}{Unit capacity cost of battery
module \(i\)}
\nomenclature[B, 11]{${\pi _{_{\rm{p}}}^{{\rm{loss}}}}$}{Unit electricity price for regulation}
\nomenclature[B, 11]{${\pi _{_{\rm{p}}}^{{\rm{Reg}}}}$}{Unit penalty price for regulation}
\nomenclature[B, 11]{ ${\pi_{p}^{{\rm{Deg}}}}$}{The penalty coefficient for battery degradation}

\nomenclature[B, 12]{$\Delta t$}{Time span for fast frequency response service }
\nomenclature[B, 13]{${\kappa _{1}}$,${\kappa _{2}}$}{The constant related to battery aging}

\nomenclature[C, 01]{${I_{bat,k,i,t}}$}{The internal current of module \(i\)}
\nomenclature[C, 02]{$P_{bat,k,i,t}$}{The internal power output of module \(i\)}
\nomenclature[C, 03]{$P_{mod,k,i,t}$}{The external power output of module \(i\)}
\nomenclature[C, 04]{$P_{k,i,t}^{loss} $}{The overall energy loss of module \(i\)}

\nomenclature[C, 04]{$P_{con,k,i,t}^{loss + }$, $P_{con,k,i,t}^{loss - }$}{The power loss of battery pack in the discharging/charging status}
\nomenclature[C, 05]{$P_{swt,k,i,t}^{loss + }$, $P_{swt,k,i,t}^{loss - }$}{The power loss of converter in the discharging/charging status}
\nomenclature[C, 06]{$P_{k,i,t}^{rr}$}{ The diode reverse loss of module \(i\)}
\nomenclature[C, 06]{$P_{k,i,t}^{\mathrm{dt}}$}{ The diode dead time loss of module \(i\)}
\nomenclature[C, 06]{$P_{k,i,t}^{\rm{COSS}}$}{ The MOSFET output capacitance loss of module \(i\)}
\nomenclature[C, 06]{$P_{k,i,t}^{\rm{G}}$}{ The gate charge loss of module \(i\)}

\nomenclature[C, 07]{$P_{{\rm{mbss}},k,t}$}{ The power output of battery system \(k\)}

\nomenclature[C, 08]{$So{C_{k,i,t}}$}{The state of charge of module \(i\)}
\nomenclature[C, 08]{$Score_{i,t}^{{\rm{ch}}}$}{The score of the prioritization for module $i$ in the charging status}
\nomenclature[C, 08]{$Score_{i,t}^{{\rm{dch}}}$}{The score of the prioritization for module $i$ in the discharging status}
\nomenclature[C, 08]{${{\upsilon_\phi }}$ ,${{\upsilon_\psi }}$}{The charging/discharging cycle depth}

\nomenclature[C, 09]{${V_{OCV,k,i,t}}$}{The open circuit voltage of battery pack in module \(i\)}
\nomenclature[C, 10]{$u_t$}{The state of charge/discharge of module \(i\)}

\nomenclature[C, 11]{\(\alpha_{k,i,t}\)}{The activation state of battery module \(i\)}
\nomenclature[C, 11]{\(\Omega\)}{The aging degree of module}
\nomenclature[C, 12]{$\eta_{k,i,t}^{\bmod}$}{The operation efficiency of module $i$}

\nomenclature[D, 01]{BESS}{Battery energy storage systems}
\nomenclature[D, 01]{ARIMA}{Auto-regressive integrated moving
average}
\nomenclature[D, 01]{MPC}{Model predictive control}
\nomenclature[D, 01]{FFR}{Fast frequency response}
\nomenclature[D, 01]{SoC}{State of charge}
\nomenclature[D, 01]{SoH}{State of health}
\nomenclature[D, 01]{EV}{Electric vehicles}
\nomenclature[D, 01]{MIQCP}{Mix-integer quadratic constrained program}
\nomenclature[D, 01]{PJM}{Pennsylvania—New Jersey—Maryland}

\printnomenclature

\section{Introduction}
  
Modular battery energy storage systems (BESSs) are composed of several independent battery packs, offering significant advantages to enhance operational efficiency~\cite{module}.
The battery packs in a battery system exhibit heterogeneous characteristics in terms of available capacity, energy conversion efficiency, and maximum power, due to different states of health (SoH) and manufacturers~\cite{mbess}. 
As an important type of flexible resource, modular BESSs can provide a variety of grid services, such as peak shaving and valley filling, energy arbitrage, voltage support, and frequency regulation~\cite{BESS}.

Globally, most electricity markets now offer the so-called fast frequency response (FFR) that requires fast response speed~\cite{FFR2020}. 
Due to the fast response and high power density of lithium-ion battery packs, grid-connected BESSs have great potential in providing FFR service and earning financial benefits according to the pay-for-performance mechanism. However, the battery systems are required to be constantly cycled in response to the signal when providing FFR, which generates energy conversion loss and accelerates battery degradation~\cite{he2024resilient}. 
Recent research has also highlighted the critical importance of degradation modeling and thermal management in ensuring the safe and efficient operation of battery systems under varied grid service demands, providing insights into optimizing performance and extending battery lifespan\cite{RL1,RL2,RL3,RL4,RL5,RL6}.
Therefore, it is necessary to carefully consider the operation characteristics of battery systems and determine the optimal response strategy for frequency support.

There are many studies focusing on the frequency regulation by grid-connected BESSs. 
Some of the literature have concentrated on the decision-making of battery systems for frequency support over long scheduling horizon and the coordination with other types of grid services.
Ref.~\cite{2022day_ahead} proposed a day-ahead optimization strategy for shared BESS to collaboratively provide Primary frequency support and FFR service, but the uncertainty of regulation signal was not considered.
The authors of Ref.~\cite{PSVFFR} explored the comprehensive value of peak shaving in long timescale and frequency support in short timescale for battery systems, and proposed a generalized coupling approach for batteries to achieve comprehensive multiplexing of the two services.
In Ref.~\cite{2018fr} and Ref.~\cite{2022multi}, the robust optimization approach was presented to overcome the uncertainty of regulation signal for the long-term scheduling of batteries toward FR, which could lead to the conservative results of performance in regulation market.

In addition to studies on battery scheduling for frequency regulation in the coarse timescale, there has been effort on the significant uncertainty of regulation signals and determining the optimal response of batteries in the fine timescale. 
Some literature formulated the optimal response policy for battery systems based on model predicted control~(MPC) approach~\cite{2018control}\cite{Packetized2023}. 
In Ref.~\cite{2018MPC}, a stochastic MPC framework was presented for 
battery systems that participated in both energy and regulation markets.
In Ref.~\cite{li2022deepL}, a deep learning-based power management method integrated with MPC framework was proposed for batteries providing regulation services. 
In addition to MPC approaches that rely on prediction accuracy, some studies have also focused on online scheduling strategies for battery systems in the regulation market.
Ref.~\cite{ma2023life} proposed a life-aware online optimization strategy based on the Lyapunov drift-plus-penalty method for batteries providing FFR in PJM market, which does not require regulation signal prediction and historical data.
Ref.~\cite{2024two_stage} further utilized a threshold policy for the regulation response of batteries which can balance the service quality and battery health.
Ref.~\cite{performance2024} considered the power throughput and temperature aging effects of batteries and proposed an aging-aware real-time scheduling strategy to improve service profitability and markedly extend battery lifespan. 

In the aforementioned studies on operational strategies of battery systems for frequency regulation, the modeling of batteries is at the system level and standardized.
Some studies accounted for the heterogeneity of storage units or subsystems in the battery systems and specifically investigate optimized power allocation strategies when offering grid services.
Ref.~\cite{2023scalable} established an electro-thermal model for scalable battery systems and proposed a clustering-based  hierarchical optimization approach to allocate power between and within battery clusters.
The authors in Ref.~\cite{status2024} considered the regulation capability of battery units and proposed the multi-state interval for each unit based on SoC, optimizing the economic benefit of battery storage clusters.
Ref.~\cite{liang2023hierarchical} explored the potential of coordinating a large number of small battery systems in distribution networks to participate in frequency regulation and proposed a hierarchical resilience enhancement strategy for battery management.
For the modeling of energy conversion efficiency, the authors in Ref.~\cite{2022efficiency} took the energy loss in converters of battery modules into consideration and proposed a real-time power allocation strategy based on offline genetic algorithm and online SoC management loop.
Ref.~\cite{liu2023eff_evaluation} proposed a regulation strategy that combines a power rolling distribution module and an efficiency evaluation module. 
This strategy optimizes the regulation power allocation between batteries and traditional power units, as well as among different units of battery systems, based on cost, revenue, and regulation performance.

In the aforementioned studies, most of the literature focused on power allocation approaches for multiple subsystems, primarily considering the differences in SoCs and power limits. 
Some studies further modeled the operation efficiency and battery aging cost.
However, these models often lack precision, neglecting the heterogeneous lifespan and dynamic efficiency of battery modules in their operation. 
Besides, the effective real-time scheduling strategies for heterogeneous battery clusters providing FFR service are lacked in the current studies.

To address these gaps, this paper proposes a novel performance-aware scheduling approach for modular battery systems towards FFR to co-optimize regulation performance and battery operation cost, which is suitable for frequency support by batteries in the fine timescale.

The main contributions of this paper are two fold:
\begin{enumerate}
    \item A performance-aware operation model towards FFR service is proposed for battery systems with multiple heterogeneous battery modules. Efficiency-aware constraints are established, which are effective for identifying energy loss caused by power conduction and switching in each battery module.
    The cycle-based battery aging model suitable for FFR is also integrated, which utilizes real-time battery cycling information to identify the aging cost of each other.

    \item A Priority Evaluation-based MPC approach is proposed for the optimal response of battery systems to the regulation signals with strong uncertainty. 
    In this method, a heuristic algorithm considering the operation efficiency and SoC balance is presented to determine the output status of each battery modules within the system during the scheduling horizon. 
    With this algorithm, the computational burden of FFR is significantly reduced.
  
\end{enumerate}


\section{Performance-Aware Operation Model For Battery Modules}
In this section, a performance-aware optimization model of battery modules towards FFR is presented, which involves the operation of an independent battery module and modular battery systems during discrete scheduling intervals in the pay-for-performance regulation market. 

\subsection{Operation Mode of Multiple Battery Modules}


In a distributed modular battery system, each battery pack connects to a DC bus through a controllable DC-DC converter, allowing independent power control. These systems link to local loads and distribution networks via a bidirectional AC-DC inverter, enabling participation in the pay-for-performance regulation market through aggregation by virtual power plants.

This study adopts the PJM regulation market policy, where the aggregator of distributed battery systems provides fast frequency response services, specifically the Dynamic Regulation Signal (RegD). Considering the operational characteristics of battery modules, the aggregator determines the optimal power distribution based on RegD signals and send commands to each system for implementing the frequency response.

\begin{figure}
\centering
\includegraphics[width=3.7in]{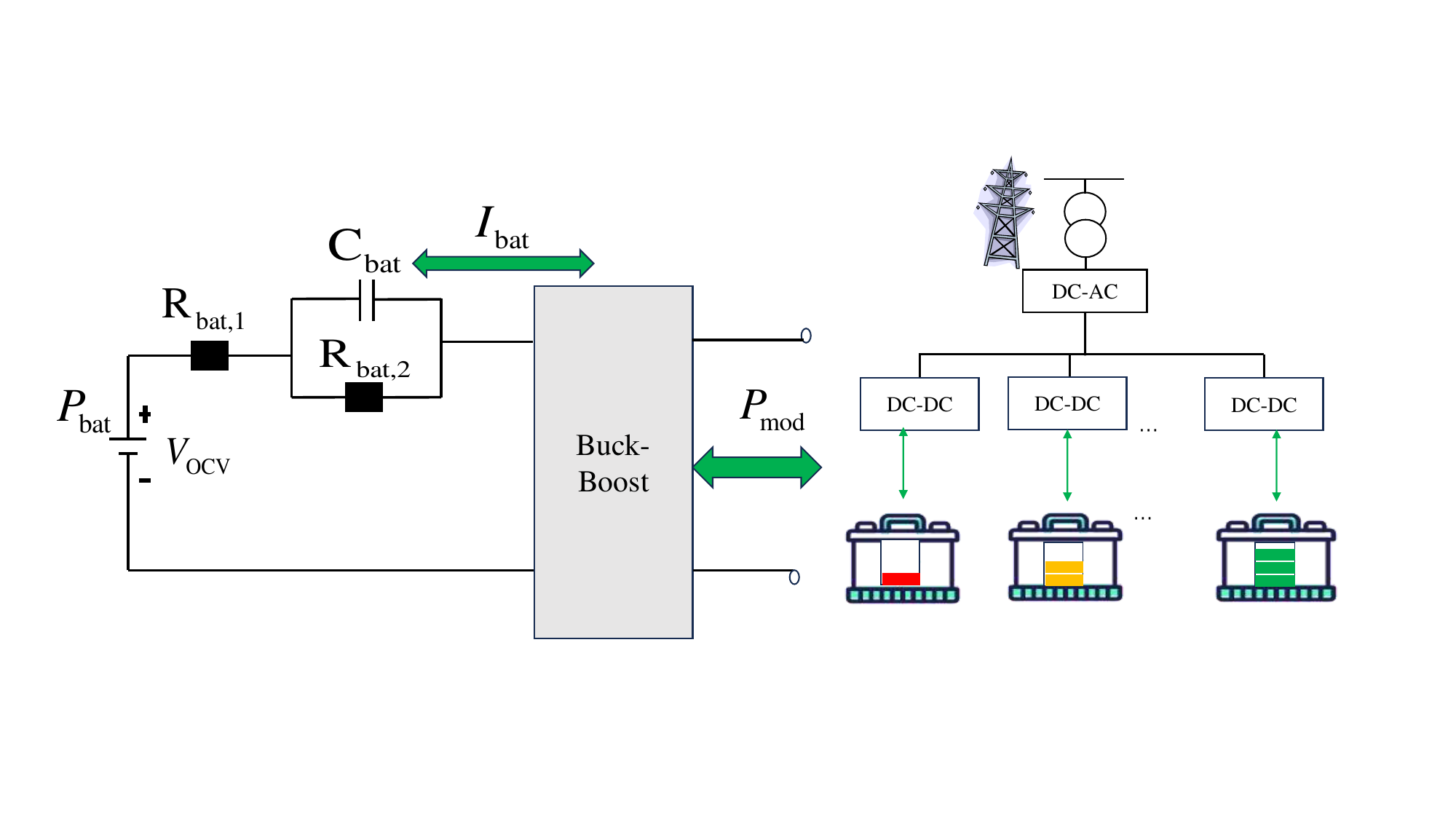}
\caption{The equivalent circuit of a battery module integrated in the BESS }
\label{fmodule}
\end{figure}

\subsection{Efficiency-Aware Constraints for Battery Modules }
In this work, we take the typical dc-side-parallel modular BESS as a reference to model the battery system~\cite{DC}, whose structure is common in electric vehicle charging and swapping stations, PV battery systems in integrated park and second-life battery systems~\cite{cui2023operation}.
The dc-side-parallel modular battery systems consists of several battery modules connected in parallel, and each pack is linked with a bidirectional half bridge buck-boost converter, as shown in Fig.~\ref{fmodule}.
To accurately identify the energy conversion efficiency of each module, it is necessary to consider the power flowing through the internal circuit and converter, as well as the conduction and switching losses. 
The modeling of energy conversion inside the module can be derived from the first order RC equivalent circuit\cite{zhang2017circuit} and the conversion efficiency of modular DC converters in Ref.~\cite{2022efficiency}.

The generalized efficiency-aware constraints for both the charging and discharging status of battery modules are presented in this subsection.

The power balance of a battery module is shown as constraint\eqref{p_balance}.
\vspace{-0.1cm} 
\begin{equation}
\begin{aligned}
\label{p_balance}
\begin{array}{l}
{P_{bat,k,i,t}} = {P_{mod,k,i,t}} + P_{con,k,i,t}^{loss + } - P_{con,k,i,t}^{loss - }\\+ P_{sw,k,i,t}^{loss + } - P_{sw,k,i,t}^{loss - },\\
\forall k \in \mathcal{K},i \in \mathcal{N},t \in \mathcal{T}
\end{array}
\end{aligned}
\end{equation}

where ${P_{bat,k,i,t}}$ and ${P_{mod,k,i,t}}$ are respectively the internal and external power output for battery module $i$ in system $k$ during time span $t$; $P_{con,k,i,t}^{loss + }$ and $P_{con,k,i,t}^{loss - }$ are respectively the conduction power loss in the discharging and charging status; $P_{sw,k,i,t}^{loss + }$ and $P_{sw,k,i,t}^{loss - }$ are respectively the switching power loss in discharging and charging status .

\vspace{-0.1cm} 

The conduction loss of a battery module can be calculated by quadratic constraint~\eqref{conduction}. 
\begin{equation}
\begin{aligned}
\label{conduction}
\begin{array}{l}
P_{con,k,i,t}^{loss + } + P_{con,k,i,t}^{loss - } \\ \ge I_{_{bat,k,i,t}}^2 \cdot (R_{{\rm{bat,}}k,i}^{} + R_{{\rm{on,}}k,i}^{} + DC{R_{k,i}}),\\  
\forall k \in \mathcal{K},i \in \mathcal{N},t \in \mathcal{T}
\end{array}
\end{aligned}
\end{equation}
where $R_{{\rm{bat,}}k,i}^{}$, $ R_{{\rm{on,}}k,i}^{}$ and $DC{R_{k,i}}$ are the internal resistance of the battery pack, the on-resistance of the MOSFET and the equivalent resistance of the inductance in the converter of battery module $i$, respectively.

The buck-boost circuit connecting the battery pack to the DC bus generates switching losses during the operation, consisting of five components, namely the voltage and current overlap loss ($P_{k,i,t}^{\rm{V- I}}$), the diode dead time loss ($P_{k,i,t}^{\rm{dt}}$), diode reverse loss ($P_{k,i,t}^{rr}$), MOSFET output capacitance loss ($P_{k,i,t}^{\rm{COSS}}$) and the gate charge loss ($P_{k,i,t}^G$). The calculation of these components are shown as follows.

\begin{equation}
\begin{aligned}
\begin{array}{l}
\label{switching}
P_{sw,k,i,t}^{loss + } + P_{sw,k,i,t}^{loss - } = P_{k,i,t}^{\mathrm{V - I}} + P_{k,i,t}^{\mathrm{dt}} + P_{k,i,t}^{rr} \\ +P_{k,i,t}^{\mathrm{COSS}} + P_{k,i,t}^G, \\
\forall  k \in \mathcal{K}, i \in \mathcal{N}, t \in \mathcal{T}
\end{array}
\end{aligned}
\end{equation}
\vspace{-0.1cm} 
\begin{equation}
\begin{aligned}
\label{mos_loss}
&\left\{ \begin{array}{l}
P_{k,i,t}^{{\rm{V - I}}} = \frac{{\left( {{V_{{\rm{dc}},k,i}} + {V_{{\rm{DS}},k,i}}} \right)}}{2} \cdot {f_{sw}} \cdot \left( {{t_r} + {t_f}} \right)\\
P_{k,i,t}^{\mathrm{dt}} = V_{\mathrm{DS},k,i} \cdot I_{\mathrm{bat},k,i,t} \cdot f_{\mathrm{sw}} \cdot \left( {t_r} + {t_f}\right)\\
P_{k,i,t}^{rr} = {Q_{{\rm{rr}},k,i}} \cdot V_{{\rm{dc}},k,i}^{} \cdot {f_{sw}}\\
P_{k,i,t}^{{\rm{COSS}}} = C_{k,i}^{{\rm{Coss}}} \cdot V_{{\rm{dc}},k,i}^2 \cdot {f_{sw}}\\
P_{k,i,t}^G = \left( {Q_{{\rm{g1}},k,i}^{} + Q_{{\rm{g2}},k,i}^{}} \right) \cdot V_{{\rm{GS}},k,i}^{} \cdot {f_{sw}}
\end{array}\right.,
\\
&\quad \forall  k \in \mathcal{K}, i \in \mathcal{N},t \in \mathcal{T}
\end{aligned}
\end{equation}

where ${V_{{\rm{dc}},k,i}}$, ${V_{{\rm{DS}},k,i}}$ and $V_{{\rm{GS}},k,i}$ are respectively the DC bus side, drain-source and gate-source voltage of battery pack $i$; ${f_{sw}}$ is the switching frequency of the MOSFET; ${{t_r}}$ and ${{t_f}}$ are the dead times during switching;  ${{t_{{\rm{dtr}}}}}$ and  ${{t_{{\rm{dtf}}}}}$ are the rising and falling time of the switching transition; 
$C_{k,i}^{{\rm{Coss}}}$ ,${Q_{{\rm{rr}},k,i}}$  are the output capacitance and the diode reverse recovery charge; $Q_{{\rm{g1}},k,i}$ and $Q_{{\rm{g2}},k,i}$ are the total gate charge of the two MOSFETs in the converter.

The safety constraint of battery modules is given as constraint~\eqref{safety}.
\begin{equation}
\begin{aligned}
\label{safety}
\begin{array}{l}
{\alpha _{k,i,t}} \cdot {I_{bat,k,i,\min }} \le {I_{bat,k,i,t}} \le {\alpha _{k,i,t}} \cdot {I_{bat,k,i,\max }}, \\
\forall  k \in \mathcal{K}, i \in \mathcal{N}, t \in \mathcal{T}
\end{array}
\end{aligned}
\end{equation}
where ${\alpha _{k,i,t}}$ is the binary variable indicating the activation state of battery module $i$ in system $k$ during time span $t$.

The relationship of open-circuit voltage, internal current and SoC within the battery module of is given by constraints~\eqref{V_I}-\eqref{V_SOC}. It is mentioned that the open circuit voltage of the battery can be approximated as a linear function of SoC in the operation range of $\left[ {So{C_{i,\min }},So{C_{i,\max }}} \right]$ (e.g. [0.1, 0.9] for lithium-ion battery pack)~\cite{2023gu}.
\vspace{-0.1cm} 
\begin{equation}
\label{V_I}
I_{bat,k,i,t} = P_{bat,k,i,t}/{V_{OCV,k,i,t}},\quad
\forall  k \in \mathcal{K}, i \in \mathcal{N},t \in \mathcal{T}
\end{equation}
\vspace{-0.3cm} 
\begin{equation}
\label{V_SOC}
{V_{OCV,k,i,t}} \approx {K_0} + {K_1}SoC_{k,i,t},\quad
\forall  k \in \mathcal{K}, i \in \mathcal{N},t \in \mathcal{T}
\end{equation}
where ${K_0}$ and ${K_1}$ denote the linear fitting parameters; $So{C_{k,i,t}}$ denotes the SoC of battery module $i$ in system $k$ during time interval $t$.

The relationship between the value of each item for power loss and the binary variable $u_t^{}$, which indicates the status of charge and discharge, is derived from the big M method~\cite{bigM}, as shown in constraint~\eqref{ut}.
\begin{equation}
\begin{aligned}
\label{ut}
\begin{array}{l}
\left\{ \begin{array}{l}
0 \le P_{con,k,i,t}^{loss + } \le {\rm{M}} \cdot u_t\\
0 \le P_{con,k,i,t}^{loss - } \le {\rm{M}} \cdot (1 - u_t)\\
0 \le P_{sw,k,i,t}^{loss + } \le {\rm{M}} \cdot u_t\\
0 \le P_{sw,k,i,t}^{loss - } \le {\rm{M}} \cdot (1 - u_t)
\end{array} \right.,\\ 
\forall  k \in \mathcal{K}, i \in \mathcal{N},t \in \mathcal{T}
\end{array}
\end{aligned}
\end{equation}

The  changing of SoC between two adjacent time intervals for battery module $i$ can be presented by constraint~\eqref{socchange}, which is relaxed through the Big M method.
\begin{equation}
\begin{aligned}
\label{socchange}
&\left\{ \begin{array}{l}
So{C_{k,i,t}} \le So{C_{k,i,t - 1}} - {P_{bat,k,i,t}} \cdot \Delta t/{E_{k,i}} +{\rm{M}} \cdot (1 - u_t)\\
So{C_{k,i,t}} \ge So{C_{k,i,t - 1}} - {P_{bat,k,i,t}} \cdot \Delta t/{E_{k,i}} -{\rm{M}} \cdot (1 - u_t)\\
So{C_{k,i,t}} \le So{C_{k,i,t}} + {P_{bat,k,i,t}} \cdot \Delta t/{E_{k,i}} +{\rm{M}} \cdot u_t\\
So{C_{k,i,t}} \le So{C_{k,i,t}} + {P_{bat,k,i,t}} \cdot \Delta t/{E_{k,i}} - {\rm{M}} \cdot u_t
\end{array} \right.,\\
&\forall  k \in \mathcal{K}, i \in \mathcal{N},t \in \mathcal{T}
\end{aligned}
\end{equation}
where $E_{k,i}$ denotes the maximum energy capacity of battery module $i$ in system $k$.

The operation range of SoC for each battery module in the battery system is limited by constraint~\eqref{SOClimit}.
\begin{equation}
\label{SOClimit}
SoC_{k,i,\min } \le SoC_{k,i,t}\le SoC_{k,i,\max },  \quad 
\forall k \in \mathcal{K}, i \in \mathcal{N},t \in \mathcal{T}
\end{equation}

\subsection{Aging Cost Modelling For Battery Modules }
Due to the frequent charge and discharge cycles required for FFR, the  charge/discharge scheduling of battery modules needs to consider the degradation cost of each module based on the cycling information.
However,  the cycles of modules during scheduling period are not continuously differentiable.
Since this work focuses on real-time scheduling of modules towards FFR in the fine timescale, the cycle-based battery aging cost model in the coarse timescale cannot be directly integrated. 
Therefore, this work adopts the subgradient of real-time aging to measure the aging cost during scheduling. 
The the relationship between the subgradients of battery aging and cycles can be derived from literature~\cite{Subgradient}.

In this work, a power function derived from literature \cite{Cycle-Based} is utilized to describe the relationship between the cycle depth and the aging degree of a battery module during one cycle, as shown by equation \eqref{cycle_aging}.
\vspace{-1mm}
\begin{equation}
\label{cycle_aging}
\Omega (\upsilon ) = {\kappa _{1}} \cdot \upsilon _d^{{\kappa _{2}}}\
\end{equation}
in which $\Omega$ is the aging degree and ${\upsilon_d}$ is the special cycle depth in one cycle.

The relationship between the number of effective cycles under a specific cycle depth and the number of cycles under the 100\% cycle depth can be expressed by the following equation~\eqref{cycle_DoD}, derived from Ref.~\cite{Life_cycle}.
\vspace{-1mm}
\begin{equation}
\label{cycle_DoD}
{\upsilon _{100\% }} \cdot N_{100\% }^{cycle} = {\upsilon_d} \cdot N_d^{cycle}
\end{equation}

With equation~\eqref{cycle_DoD}, the value of ${\kappa _1}$ for each module can be derived as follows.
\vspace{-1mm}
\begin{equation}
\label{k1_va}
\left\{ \begin{array}{l}
\frac{1}{{N_d^{cycle}}} = \frac{{{\upsilon_d}}}{{{\upsilon_{100\% }} \cdot N_{100\% }^{cycle}}}\\
{\kappa _1} = \frac{1}{{{\upsilon_{100\% }} \cdot N_{100\% }^{cycle}}}
\end{array} \right.
\end{equation}

The accumulated aging cost of a battery module after 
T scheduling periods can be calculated by equation~\eqref{aging_cost}, derived from Ref.~\cite{2018cycle}.
\vspace{-1mm}
\begin{equation}
\label{aging_cost}
\begin{array}{l}
{F_{k,i}} = {E_{k,i}}\pi _{k,i}^{{\rm{bat\_cost}}}(\sum\limits_{\phi  \in {\Phi _t}} {{\kappa _\phi }} \Omega \left( {{\upsilon_\phi }} \right) + \sum\limits_{\psi  \in {\Psi _t}} {{\kappa _\psi }} \Omega \left( {{\upsilon_\psi }} \right))
\end{array}
\end{equation}
in which the $\pi _{k,i}^{{\rm{bat\_cost}}}$ is the unit capacity cost of battery modules, ${{\upsilon_\phi }}$ and ${{\upsilon_\psi }}$ are respectively the charging cycle depth and discharging cycle depth in a cycle.

When calculating the real-time subgradients of battery aging, it is necessary to formulate a sequence of extremum points$\{ {a_1},{a_{j - 2}},...,{a_{j - 1}},{a_j}\}$ for the SoC curve of each module.
Each extremum point represents the point at which the charging or discharging status changes.
The initial SoC of the module in the latest real-time scheduling period is always regarded to be at the last extremum point ${a_{j}}$.

It is mentioned that only the power optimization of all modules for FFR service during real-time period is focused in this work. 
Consequently, when the RegD curve crosses the zero point, the SoC extremum points of all modules will be updated.

When the module is in the charging state during the previous period, and still the charging command is received during the current period, the charge at period $t$ can be considered as part of the charging cycle from the previous period $t-1$.
Thus, the absolute difference between the last  two extremum points in the extremum point sequence represents the cycle depth of the current charging cycle.

The relationship between the current charging cycle depth and the real-time charging power at time $t$ of one module can be indicated by the following constraint~\eqref{charging_DoD}.

\begin{equation}
\begin{aligned}
\label{charging_DoD}
\begin{array}{l}
{\upsilon_{\phi,k,i,t} } = \left| {SoC_{k,i,{t-1}} - {a_{k,i,j - 1}}} \right| + \frac{P_{bat,k,i,t}^{}}{E_{k,i}} \Delta t,\\
\forall  k \in \mathcal{K}, i \in \mathcal{N}, t \in \mathcal{T}
\end{array}
\end{aligned}
\end{equation}

The real-time aging subgradient of module $i$ in system $k$ with respect to the charging power in the current period $t$ can be calculated by constraint ~\eqref{charge_sg}.
\begin{equation}
\label{charge_sg}
\begin{aligned}
&\frac{\partial F\left( P_{bat,k,i,t} \right)}{\partial P_{bat,k,i,t}^{ch}} \\
&= \frac{1}{2} \Delta t \pi_{k,i}^{\text{bat\_cost}} \Omega' \left( \left| SoC_{k,i,{t-1}} - a_{k,i,j - 1} \right| + \frac{P_{bat,k,i,t}^{ch}}{E_{k,i}} \Delta t \right) \\
&\approx \frac{1}{2} \Delta t \pi_{k,i}^{\text{bat\_cost}} \Omega' \left( \left| SoC_{k,i,{t-1}} - a_{k,i,j - 1} \right| \right),\\
&\forall  k \in \mathcal{K}, i \in \mathcal{N}, t \in \mathcal{T}
\end{aligned}
\end{equation}

Similarly, when the module is in a discharging state in the previous period $t-1$ and the discharging command is also received in the current period $t$, the calculation of the real-time cycle depth and the aging subgradient is the same as in the charging status, also corresponding to constraints~\eqref{charging_DoD}-\eqref{charge_sg}.

If the current RegD command switches directly between charging and discharging,
a new extremum point will be generated, initiating the formation of an unstable charging/discharging half-cycle. 
At this interval $t$, the cycle depth of the unstable charging/discharging half-cycle can be considered infinitesimally small, as shown by constraint~\eqref{infinite}.

\begin{equation}
\begin{aligned}
\label{infinite}
\begin{array}{l}
\frac{{\partial F\left( {{P_{bat,k,i,t}}} \right)}}{{\partial P_{_{bat,k,i,t}}^{}}} = \Delta t\pi _{k,i}^{{\rm{bat\_cost}}}\frac{1}{2}\xi,\\
\forall  k \in \mathcal{K}, i \in \mathcal{N}, t \in \mathcal{T}
\end{array}
\end{aligned}
\end{equation}

in which $\xi$ is a very small constant.

Apparently, constraints~\eqref{charge_sg} are nonlinear and need to be linearized.
For the absolute function in constraint~\eqref{charge_sg}, it can be relaxed by constraint~\eqref{abs}, in which ${\omega_{k,i,t}}$ is the substitute of the absolute value.
\begin{equation}
\begin{split}
\label{abs}
\left\{ \begin{array}{l}
{\omega _{k,i,t}} \ge So{C_{k,i,t - 1}} - {a_{k,i,j - 1}}\\
{\omega _{k,i,t}} \ge {a_{k,i,j - 1}} - So{C_{k,i,t - 1}}
\end{array} \right.,
\forall  k \in \mathcal{K}, t \in \mathcal{T}
\end{split}
\end{equation}

For the power function involved in constraint~\eqref{charge_sg}, using piecewise linearization would introduce numbers of binary variables, significantly increasing the computational burden in the optimization problem. 
Therefore, the Taylor series approximation method can be an alternative.

For each scheduling horizon in the fine timescale, the Taylor expansion is performed at the point $\upsilon_{k,i,t}^0 = \left| {SoC_{k,i,{t_0}} - a_{k,i,j - 1}^{}} \right|$, where ${t_0}$ denotes the initial time of the horizon. 
The first-order Taylor expansion is shown by the following constraint~\eqref{taylor}.

\begin{equation}
\label{taylor}
\begin{aligned}
&\Omega '\left(  {{\omega _{k,i,t}}}  \right) \\
&= \kappa _{1,k,i} \left( \upsilon_{k,i,t}^0 \right)^{\kappa _{2,k,i}-1} \\
&\approx \kappa _{1,k,i} \left( (\upsilon_{k,i,t}^0)^{{\kappa _{2,k,i}} - 1} + (\kappa _{2,k,i} - 1) \cdot (\upsilon_{k,i,t}^0)^{{\kappa _{2,k,i}} - 2} \right. \\
&\quad \left. \cdot ({{\omega _{k,i,t}}} - \upsilon_{k,i,t}^0) \right)\\
&\forall  k \in \mathcal{K}, i \in \mathcal{N}, t \in \mathcal{T}
\end{aligned}
\end{equation}

\subsection{Optimized Operation of Battery Modules Towards FFR}
In terms of the optimized operation of BESSs participating in frequency regulation market, the power output of all battery modules in the system can be integrated in response to the real-time RegD signals.

The constraint of power balance in the battery system is given by constraint \eqref{total}, in which the ${P_{{\rm{mbss}},k,t}}$ represents the power output of battery system $k$ during time span $t$.
\begin{equation}
\label{total}
\sum\limits_i^{\rm N} {P_{{\rm{mod}},k,i,t}} = {P_{{\rm{mbss}},k,t}},  \quad 
\forall k \in \mathcal{K}, t \in \mathcal{T}
\end{equation}

The relationship between the power demand of FFR for multiple BESSs in the distribution network and RegD signals issued by PJM is given by constraint~\eqref{rt}, in which ${r_t}$ represents the value of RegD signal and $C_{{\rm{bess}}}^{{\rm{bid}}}$ represents the bidding capacity of aggregated battery systems for FFR services.
The relationship between the binary variable $u_t^{}$ and the value of regulation command $P_{\rm{t}}^{{\rm{RegD}}}$ is given by constraint \eqref{command}, in which
$u_t^{} \in \left\{ {0,1} \right\}$ represents the charging/discharging commend for the batteries.

\begin{equation}
\begin{aligned}
\label{rt}
\begin{array}{l}
P_{\rm{t}}^{{\rm{RegD}}} = C_{{\rm{bess}}}^{{\rm{bid}}} \cdot {r_t},\quad
\forall t \in \mathcal{T}
\end{array}
\end{aligned}
\end{equation}

\vspace{-0.1cm} 
\begin{equation}
\begin{aligned}
\label{command}
\begin{array}{l}
\left\{ \begin{array}{l}
P_{\rm{t}}^{{\rm{RegD}}} \ge  - {\rm{M}} \cdot (1 - u_t^{})\\
P_{\rm{t}}^{{\rm{RegD}}} \le {\rm{M}} \cdot u_t^{}
\end{array} \right.,\quad \forall t \in \mathcal{T}
\end{array}
\end{aligned}
\end{equation}

Due to the heterogeneity among the battery modules in terms of available power, energy conversion efficiency and SoC, the power allocation for each module during scheduling required to be optimized to ensure the minimum operation cost of battery systems participating in FFR.
The total energy conversion loss of each module in the system can be denoted by constraint~\eqref{totalloss}.

\begin{equation}
\begin{aligned}
\label{totalloss}
\begin{array}{l}
P_{k,i,t}^{loss} = P_{{\rm{con}},k,i,t}^{loss + } + P_{{\rm{con}},k,i,t}^{loss - } + P_{{\rm{swt}},k,i,t}^{loss + } + P_{{\rm{swt}},k,i,t}^{loss - }, \\ 
\forall k \in \mathcal{K}, i \in \mathcal{N},t \in \mathcal{T}
\end{array}
\end{aligned}
\end{equation}

Therefore, the objective function, which aims at minimizing the penalty cost of response mismatch, the overall energy loss cost and the battery aging cost during the scheduling horizon, is shown as objective function~\eqref{obj}.

\begin{equation}
\label{obj}
\begin{aligned}
\text{Minimize} \quad & \sum_{k = 1}^K \sum_{i = 1}^N \sum_{t = 1}^T 
\Bigg( \pi_{p}^{\text{loss}} P_{k,i,t}^{\text{loss}} \cdot \Delta t \\
& + \pi_{p}^{\text{Reg}} \left( \left| P_{t}^{\text{RegD}} \right| - P_{mbss,k,t} \right) \cdot \Delta t \\
& + \pi_{p}^{\text{Deg}} \frac{\partial F\left( P_{bat,k,i,t} \right)}{\partial P_{bat,k,i,t}} \Bigg)
\end{aligned}
\end{equation}

  The first term of the objective function represents the total energy loss cost due to power conduction and switching of all modules, in which the ${\pi _{p}^{{\rm{loss}}}}$ is the unit electricity price for regulation. 
  The second term denotes the penalty cost caused by response mismatch referring to the real RegD signals, in which the  ${\pi _{p}^{{\rm{Reg}}}}$ is the unit penalty price.
  The last term indicates the aging cost caused by charging/discharging cycles of all modules, in which the ${\pi_{p}^{{\rm{Deg}}}}$ is the penalty coefficient for battery degradation of modules.

 Consequently, the performance-aware operation model for multiple battery modules consists of the objective function \eqref{obj} and operation constraints \eqref{p_balance}-\eqref{totalloss}. 
 The optimization model will be further integrated into the MPC approach presented in Section~III.

\section{Priority Evaluation-based MPC Approach for FFR Services}

In this section, a priority Evaluation-based MPC method for distributed battery systems is proposed to optimize the power allocation between battery modules and rapidly implement the response to the real-time RegD signals.

\subsection{RegD Signal Prediction}

We apply the ARIMA (auto-regressive integrated moving average) model \cite{ARIMA2017} for predicting the frequency regulation signals. 
Generally, the historical data of  RegD signals is accessible, making it feasible to achieve ultra-short-term prediction of future RegD signals using the series of updated historical data. 
The predicted RegD signals can then be input into the optimization model for the most recent scheduling time interval.



\subsection{Activation Priority Algorithm for Battery Modules}

In PJM regulation market, RegD signals are dispatched every two seconds. Accurately tracking these signals for optimal power allocation across battery modules requires solving the MIQCP problem outlined in Section II with each new signal. However, as the scale of battery modules increases, the number of integer variables also rises significantly, making it difficult to solve within the two-second window using commercial solvers.

To bridge the gap between solving speed and time resolution, this paper proposes an activation priority algorithm integrated into the MPC approach. This algorithm leverages offline optimization results to fix the activation states of individual battery modules over the prediction horizon, transforming the original MIQCP problem into a QCP problem. This simplification reduces computational burden and enables approximate optimal solutions within the fine timescale of FFR.

\subsubsection{Offline Optimization}
Firstly, a certain step is chosen to traverse all possible operating points in the ${r_t} \in [ - 1,1]$, and power allocation optimization for battery modules is implemented sequentially at each operating point. 
This is to solve the original MIQCP problem with a determined RegD signal, and thus obtain the number of activated modules at different operating points. 
A lookup table is further generated to present this relationship for the application in real-time scheduling approach towards FFR.

The formulation of offline optimization is given by \eqref{obj_ol}-\eqref{con_ol}.
\begin{equation}
\label{obj_ol}
\begin{aligned}
\text{Minimize} 
\begin{array}{l}
\sum\limits_{k = 1}^K {\sum\limits_{i = 1}^N ( } \pi _p^{{\rm{loss}}}P_{k,i,t}^{{\rm{loss}}} \cdot \Delta t + \\
\pi _p^{{\rm{Reg}}}\left( {{\left| {P_{{r_t}}^{{\rm{RegD}}}} \right|} - {P_{mbss,k,t}}} \right) \cdot \Delta t)
\end{array}
\end{aligned}
\end{equation}

\begin{equation}
\begin{aligned}
\label{con_ol}
\begin{array}{l}
{\rm{s}}{\rm{.t}}{\rm{.}}\begin{array}{*{20}{c}}
{}&{(1) - (10),\begin{array}{*{20}{c}}
{(20) - (23)}&{}
\end{array}}
\end{array}
\end{array}
\end{aligned}
\end{equation}

In this problem, the constraints related to the aging subgradient do not need to be taken into account because the optimization at special operation point does not involve incremental aging cost.

\subsubsection{Online Implementation}
In response to the real-time updating RegD signals, an evaluation model of operation efficiency and SoC balance is required in order to prioritize the activation of battery modules with high energy conversion efficiency and strong SoC recovery need within the battery systems. 
For the SoC and efficiency of modules in each scheduling time interval, three sets are defined, which are $SOC = \left\{ {So{C_{1,t}},So{C_{2,t}},...,So{C_{n,t}}} \right\}$, $\eta _{bat}^{{\rm{ch}}} = \left\{ {\eta _{1,t}^{{\rm{ch}}},\eta _{2,t}^{{\rm{ch}}},...,\eta _{n,t}^{{\rm{ch}}}} \right\}$ and $\eta _{bat}^{{\rm{dch}}} = \left\{ {\eta _{1,t}^{{\rm{dch}}},\eta _{2,t}^{{\rm{dch}}},...,\eta _{n,t}^{{\rm{dch}}}} \right\}$.


The total score of the prioritization for battery modules in the charging state is calculated as equations~\eqref{chscore} - \eqref{chscore1}. 
\vspace{-1mm}
\begin{equation}
\label{chscore}
\left\{ \begin{array}{l}
Score_{i,t}^{{\rm{ch\_eff}}} = \frac{{\eta _{i,t}^{{\rm{ch}}} - \min (\eta _{}^{{\rm{ch}}})}}{{\max (\eta _{}^{{\rm{ch}}}) - \min (\eta _{}^{{\rm{ch}}})}}\\
Score_{i,t}^{{\rm{ch\_SOC}}} = \frac{{\max \left( {SOC} \right) - SoC_{i,t}^{}}}{{\max \left( {SOC} \right) - \min \left( {SOC} \right)}}
\end{array} \right.
\end{equation}
\vspace{-0.1cm} 
\begin{equation}
\label{chscore1}
\begin{array}{l}
Score_{i,t}^{{\rm{ch}}} = \omega _{eff}^{} \cdot Score_{i,t}^{{\rm{ch\_eff}}} + \omega _{SOC}\cdot Score_{i,t}^{{\rm{ch\_SoC}}}
\end{array} 
\end{equation}

Likewise, the total score of the prioritization for battery modules in the discharging state is calculated as equations~\eqref{dchscore} - \eqref{dchscore1}.  

\begin{equation}
\label{dchscore}
\left\{ \begin{array}{l}
Score_{i,t}^{{\rm{dch\_eff}}} = \frac{{\eta _{i,t}^{{\rm{dch}}} - \min (\eta _{}^{{\rm{dch}}})}}{{\max (\eta _{}^{{\rm{ch}}}) - \min (\eta _{}^{{\rm{ch}}})}}\\
Score_{i,t}^{{\rm{dch\_SoC}}} = \frac{{SoC_{i,t}^{} - \min \left( {SOC} \right)}}{{\max \left( {SOC} \right) - \min \left( {SOC} \right)}}
\end{array} \right.
\end{equation}

\vspace{-0.1cm} 
\begin{equation}
\label{dchscore1}
\begin{array}{l}
Score_{i,t}^{{\rm{dch}}} = \omega _{eff}^{} \cdot Score_{i,t}^{{\rm{dch\_eff}}} + \omega _{soc}^{} \cdot Score_{i,t}^{{\rm{dch\_SoC}}}
\end{array} 
\end{equation}

For the selection of weight coefficients in \eqref{chscore1} and \eqref{dchscore1}, 
the approach of evenly distributed weighting is adopted, which can ensure that each factor is considered equally in the priority evaluation.

\subsection{MPC-based Scheduling Strategy for FFR}



\begin{figure*}[!t]
\centering
\includegraphics[width=1\textwidth]{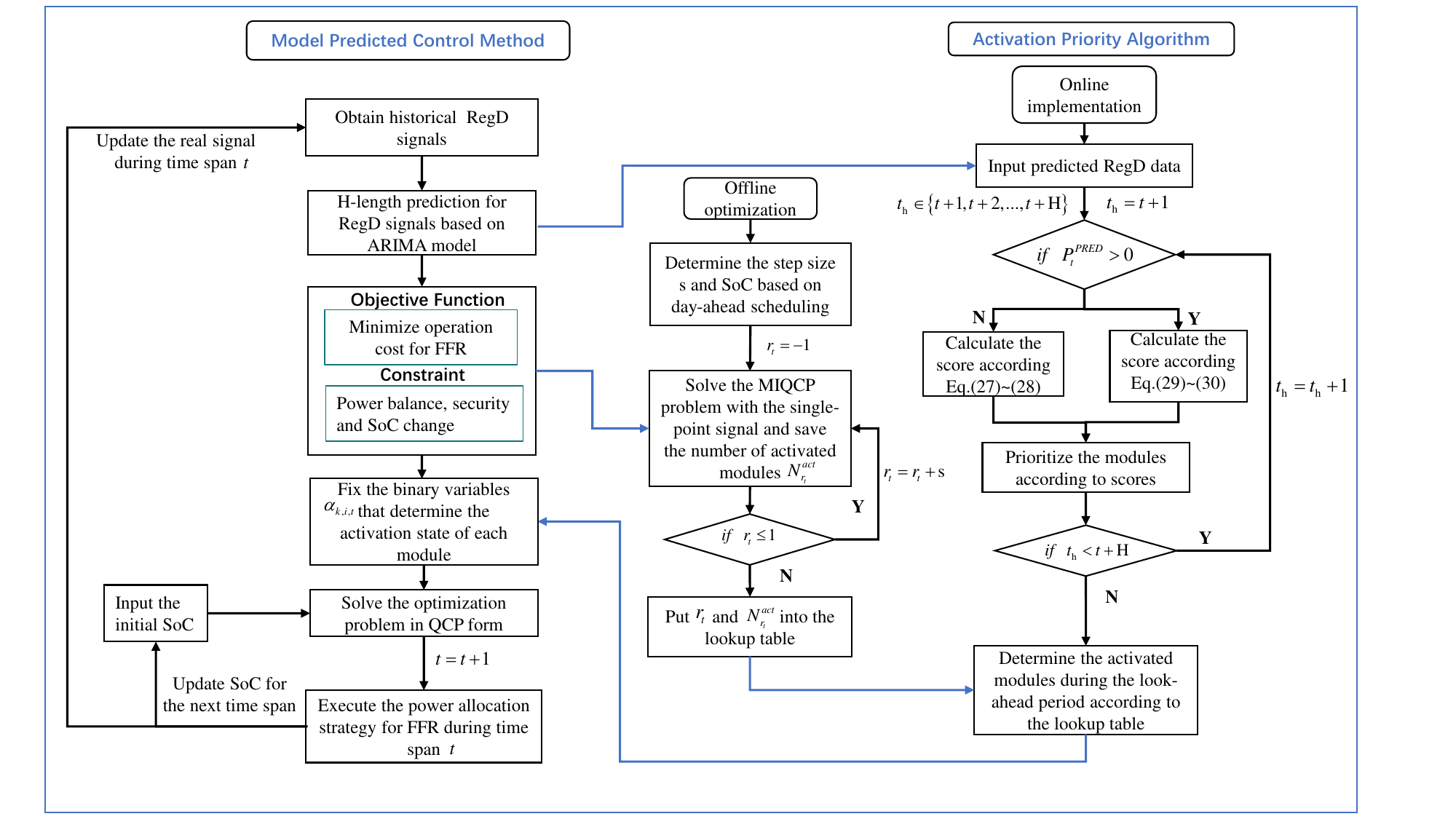}  
\caption{The priority evaluation-based MPC approach of multiple battery modules for FFR}
\label{mpc}
\end{figure*}

The MPC-based scheduling strategy consists of two parts: MPC method and the activation priority algorithm, as shown in Fig.~\ref{mpc}.

In the MPC part, an ultra-short-term prediction of future H-length RegD data is conducted using ARIMA based on historical RegD data of a certain length. 
The predicted data are then inputted into the activation priority algorithm. 
The prioritization of battery modules to participate in FFR is evaluated based on the charging or discharging command reflected by the signal and sorted according to the score. 
Utilizing the predicted data, the number of activated modules corresponding to the signal is determined by searching through the lookup table generated from offline optimization results. Consequently, the sequence of selected battery modules for FFR during the latest time interval is determined.

Based on the sequence of activated modules, the binary variables in constraint~\eqref{safety} that determine the activation state of battery modules can be fixed, i.e.,the original MIQCP problem is transformed into a QCP problem.
By solving this QCP problem, the optimal power allocation can be quickly obtained, enabling rapid implementation of frequency response during the latest scheduling interval.

\vspace{-5mm}

\section{Case Study}
We conduct case studies based on one modular BESS and multiple BESSs in regional distribution networks. 
The MPC framework is implemented using Python 3.9.2 and the optimization problem is solved by Gurobi v10.0.2 with the Pyomo environment \cite{pyomo} on a computer with an Intel Core i7-1260P 2.10 GHz CPU and 16 GB RAM.

\subsection{Case I: A modular battery system}
\subsubsection{Test System Description}
In this case, we assume that there is a single lithium-ion battery system participating in PJM frequency regulation market.
In the part of online implementation, the RegD signal data of one day from PJM market\cite{PJM_data} is utilized to simulate the operation of the battery system towards FFR service. 
The real-time RegD signals are updated every two seconds for the service.
Therefore, in this case we set the time interval of MPC-based scheduling framework to $\Delta t = 2{\rm{s}}$ and the look-ahead length is set to ${\rm{H}} = 15\Delta t$.

For the specification of the battery system in this case, we take the characteristics of battery modules in a real-field and gird-scale  heterogeneous battery system, i.e., the M5BAT project \cite{m5bat} conduct by RWTH Aachen University in Aachen, Germany for our case studies.

The nominal power and the energy capacity of the system is 6MW/ 7.5MWh, which consists of 10 storage modules utilizing five distinct battery technologies, including two types of lead-acid and three types of Li-ion batteries.
In this case, since we focus on the operation of the DC-side parallel modular battery system, it is assumed that each unit consists of a battery pack with a DC-DC converter connected to the DC bus of a microgrid.



For the operation settings of the battery system, given the fast response of the batteries, we assume that the delay time for battery modules to respond to the RegD signal is negligible.
The SoC range for each module is set between [0.1, 0.9], and the maximum charging/discharging current is set to the nominal current.
The initial SoC of each module is obtained from the day-ahead scheduling of battery aggregator toward FFR service, during which the SoC is determined according to the hourly scheduling strategy.
Since the focus of this work is not on the day-ahead scheduling stage, the introduction of day-ahead scheduling strategy is neglected in this part.

In terms of the performance evaluation, the total operation efficiencies of battery systems during scheduling can be calculated by equations~\eqref{eff_calculation} - \eqref{total_eff}.
\vspace{-0.1cm}
\begin{equation}
\begin{aligned}
\label{eff_calculation}
\left\{
\begin{array}{ll}
\eta_{k,i,t}^{\bmod} &= \frac{P_{\bmod,k,i,t}}{P_{{\rm bat},k,i,t}} \cdot u_t 
+ \frac{P_{{\rm bat},k,i,t}}{P_{{\rm mod},k,i,t}} \cdot (1 - u_t)\\
{\nu _{k,i,t}} &= \frac{P_{\bmod,k,i,t}}{P_{{\rm mbess},k,t}}
\end{array}
\right.,\\
\forall k \in \mathcal{K}, i \in \mathcal{N},t \in \mathcal{T}
\end{aligned}
\end{equation}

\begin{equation}
\label{total_eff}
\eta _{k,t}^{\text{BESS}} = \sum\limits_{i=1}^{\rm{N}} \eta _{k,i,t} \cdot \nu _{k,i,t}, \quad 
\forall k \in \mathcal{K}, \, t \in \mathcal{T}
\end{equation}

where $\eta _{k,i,t}^{\bmod }$ represents the operation efficiency of module~$i$ during time span~$t$ and ${{\omega _{k,i,t}}}$ denotes the weight of the operation efficiency of module~$i$ in the total efficiency $\eta _{k,t}^{BESS}$. 

The SoC balance degree of battery modules within the storage system during scheduling can be identified by equation~\eqref{soc_ba}.
\begin{equation}
\label{soc_ba}
D_{k,t}^{\text{SoC}} = \sum\limits_{i=1}^{\rm{N}} {{{\left| {SoC_{k,i,t}^{} - SoC_{k,t}^{avg}} \right|} \mathord{\left/
 {\vphantom {{\left| {SoC_{k,i,t}^{} - SoC_{k,t}^{avg}} \right|} {\rm{N}}}} \right.
 \kern-\nulldelimiterspace} {\rm{N}}}},\quad 
 \forall k \in \mathcal{K},t \in \mathcal{T}
\end{equation}
where ${SoC_{k,t}^{avg}}$ is the average SoC value of modules in battery system $k$ during time interval $t$.

\begin{table*}
\renewcommand{\arraystretch}{0.9}
\centering
\caption{The basic parameter of the BESS and initial condition in Case~I}
\label{tab:bess1}
\begin{tabular}{lp{2.5cm}p{1.8cm}p{1.8cm}p{1.6cm}p{1.8cm}p{1.4cm}}

\Xhline{1.5pt}
Battery Technology & Nominal power / \newline Nominal energy &  Nominal cycle \newline number &  Nominal cell \newline voltage (V) &  Nominal Current~(A)/\newline Current rate  &  Operation \newline efficiency (\%) & Initial SoC (\%)\\
\midrule
1: Pb1 & 630 kW / 1066 kWh & 1500 & 2.0 & 1025/0.59 & 74.66 & 58.48\\
2: Pb2 & 630 kW / 1066 kWh & 1500 & 2.0 & 1025/0.59 & 78.86 & 65.17\\\
3: Pb3 & 630 kW / 843 kWh & 2400 & 2.0 & 1022/0.75 & 90.32 & 50.46\\
4: Pb4 & 522 kW / 740 kWh & 2400 & 2.0 & 847/0.71 & 82.56 & 42.38\\
5: LMO1 & 630 kW / 774 kWh & 6000 & 3.7 & 886/0.81 & 97.15 & 61.65 \\
6: LMO2 & 630 kW / 774 kWh & 6000 & 3.7 & 886/0.81 & 97.02 & 52.33\\
7: LMO3 & 630 kW / 774 kWh & 6000 & 3.7 & 886/0.81 & 96.93 & 45.37 \\
8: LMO4 & 630 kW / 774 kWh & 6000 & 3.7 & 886/0.81 & 96.85 & 63.96\\
9: LFP & 630 kW / 738 kWh & 5000 & 3.2 & 820/0.85 & 95.25 & 43.78\\
10: LTO & 630 kW / 230 kWh & 12000 & 2.3 & 877/2.74 & 94.27 & 52.65\\
\Xhline{1.5pt}
\end{tabular}
\end{table*}

\subsubsection{Simulation Results}
\begin{figure}[htbp] 
    \centering
    \subfigure[Charging status]{
        \begin{minipage}[t]{\linewidth} 
            \centering
            \includegraphics[width=2.2in]{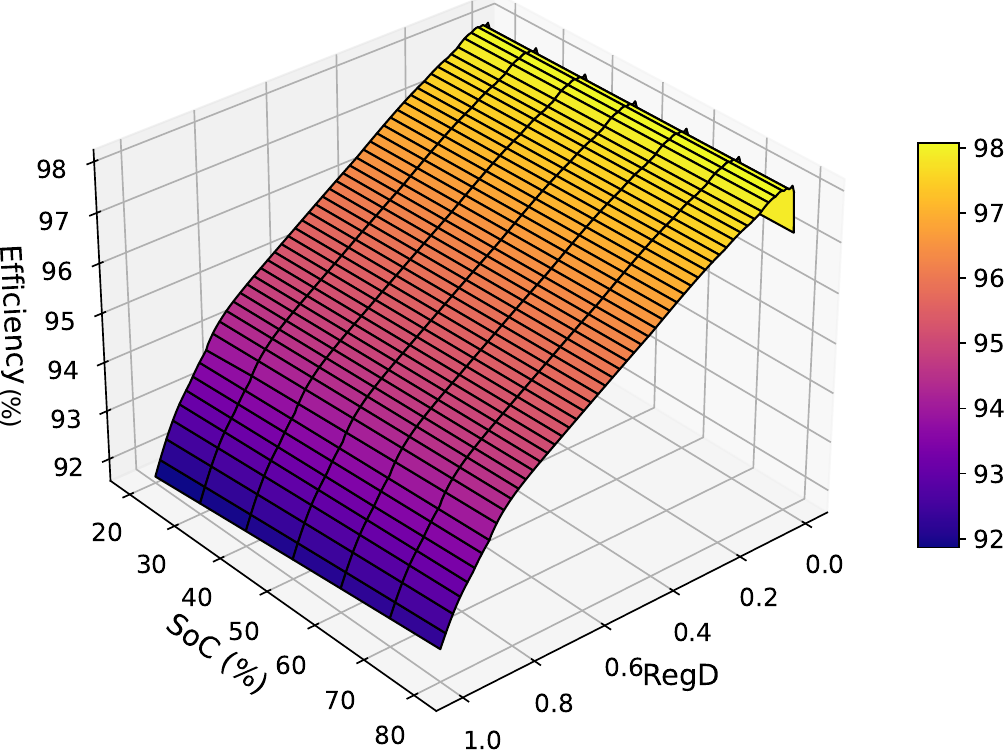} 
        \end{minipage}
    } 
    \vspace{-0.2cm} 
    \subfigure[Discharging status]{
        \begin{minipage}[t]{\linewidth} 
            \centering
            \includegraphics[width=2.2in]{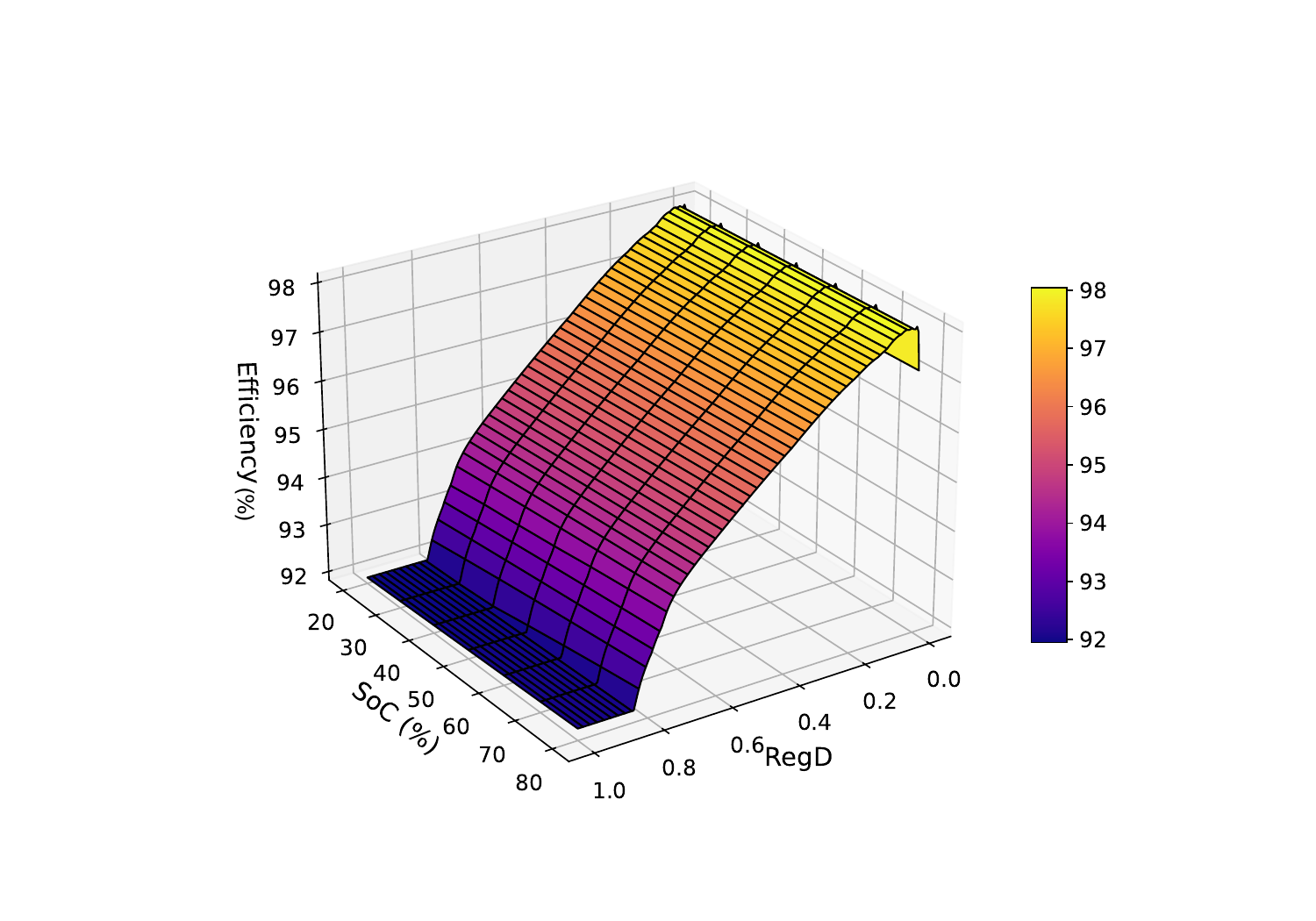} 
        \end{minipage}
    }
    \caption{The charging/discharging efficiency curve of the modular battery system in Case I.}
    \label{efficiency}
\end{figure}

\begin{figure}
    \centering
    \setlength{\abovecaptionskip}{0pt}
    \setlength{\belowcaptionskip}{0pt}
    \setlength{\floatsep}{0pt}
    \setlength{\textfloatsep}{0pt}
    \setlength{\intextsep}{0pt}

    \subfigure[]{
        \begin{minipage}[t]{0.5\linewidth}
            \centering
            \hspace*{-0.8in} 
            \includegraphics[width=3.2in]{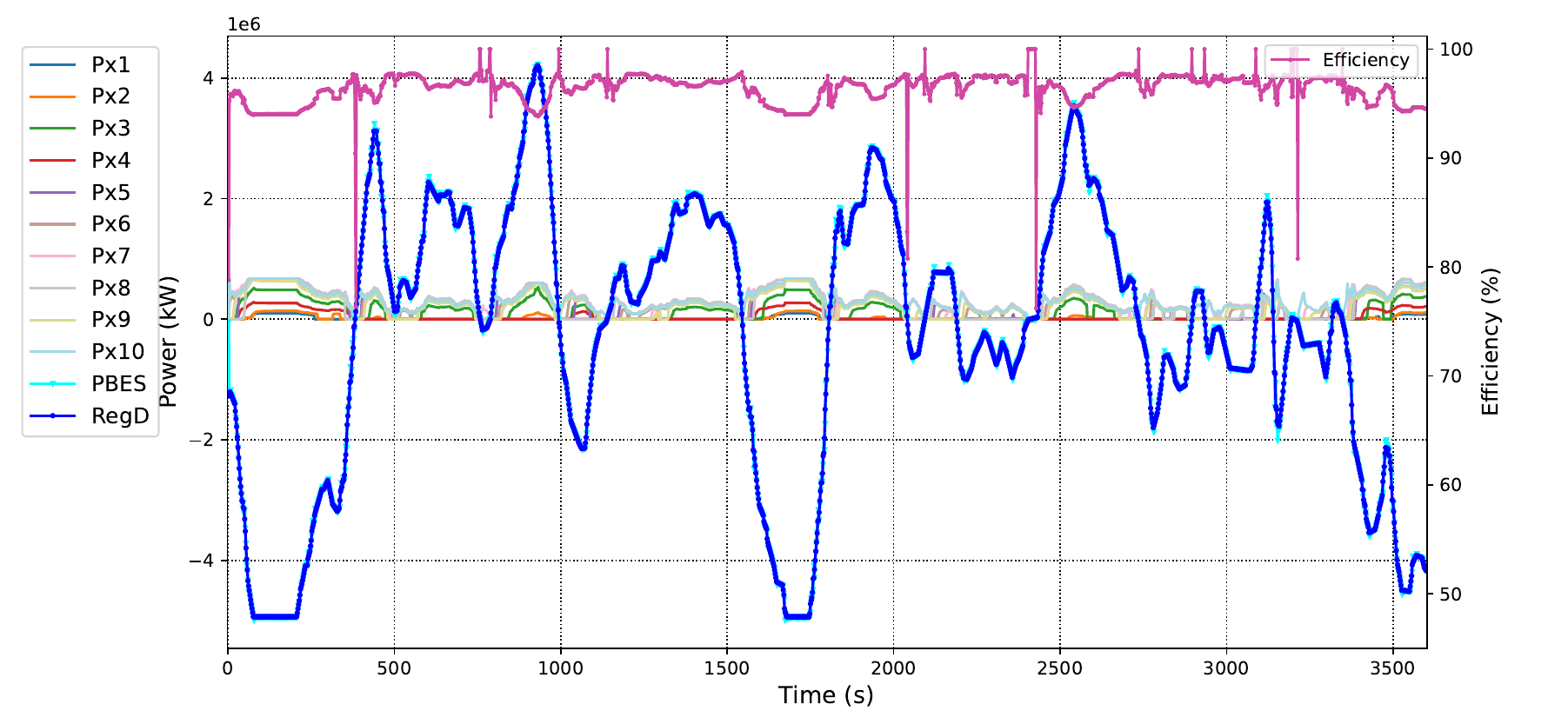} 
            \vspace{-6mm} 
        \end{minipage}
    }%
    \vspace{-3mm} 
    \\
    \subfigure[]{
        \begin{minipage}[t]{0.5\linewidth}
            \centering
            \hspace*{-0.8in} 
            \includegraphics[width=3.2in]{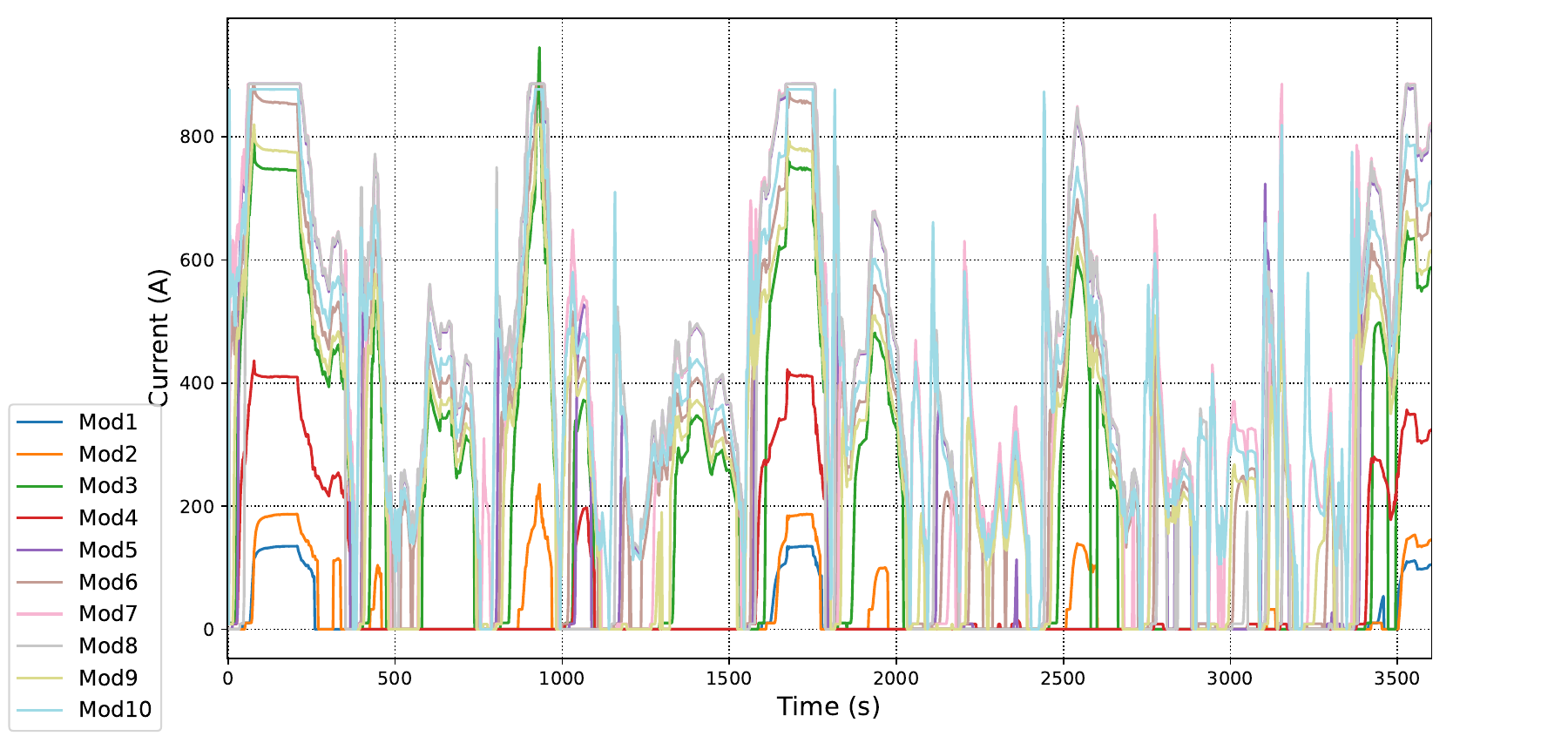} 
            \vspace{-6mm} 
        \end{minipage}
    }%
    \vspace{-3mm} 
    \\    
    \subfigure[]{
        \begin{minipage}[t]{0.5\linewidth}
            \centering
            \hspace*{-0.8in} 
            \includegraphics[width=3.2in]{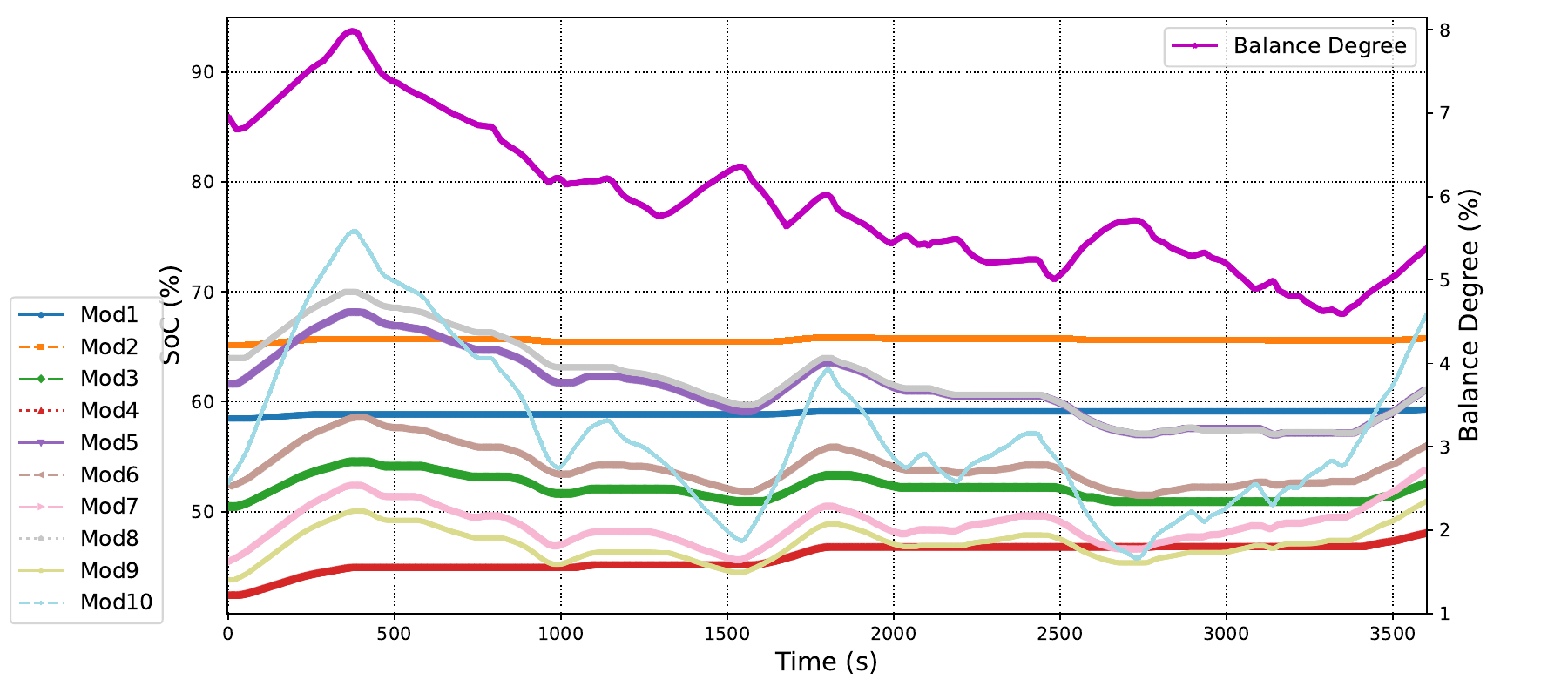} 
            \vspace{-6mm} 
        \end{minipage}
    }%
    \vspace{-3mm} 
    \\
    \subfigure[]{
        \begin{minipage}[t]{0.5\linewidth}
            \centering
            \hspace*{-0.8in} 
            \includegraphics[width=3.2in]{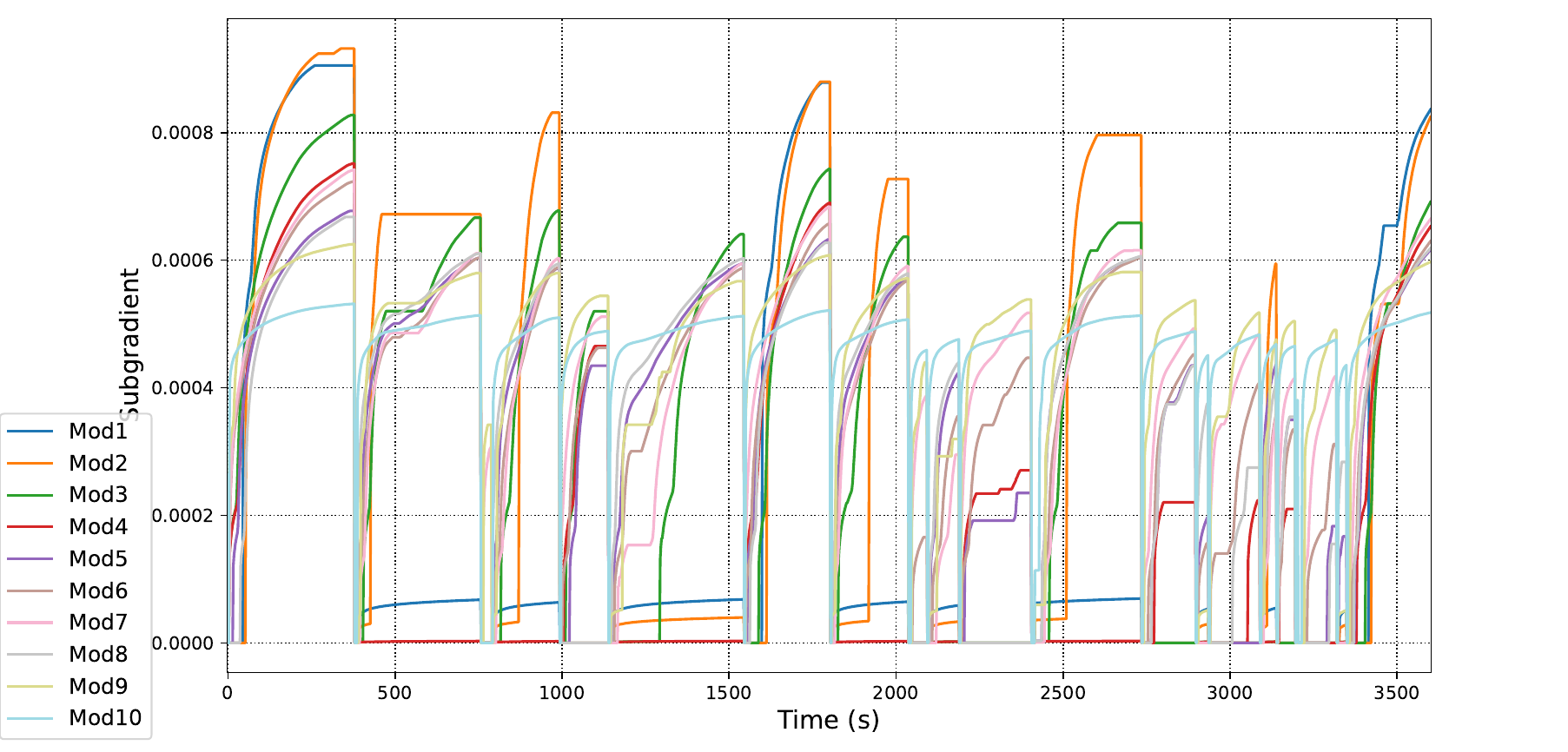} 
        \end{minipage}
    }%
    
    \caption{The scheduling results of multiple battery systems in Case II.
    (a)~The power curve and the operation efficiency curve. (b)~The curve of internal current. (c)~The SoC curve and the SoC balance degree curve. (d)~The aging subgradient curve.}
    \label{m5case}
\end{figure}

The offline optimization proposed in Section III.B. is first performed according to the step size of ${\rm{s}} = 0.005$ to obtain the lookup table for the real-time scheduling framework. 
The overall charging/discharging efficiency curve of the modular battery system is also obtained from the optimization results as shown in Fig.~\ref{efficiency}.
By simulating the real-time scheduling for RegD service in PJM market from 18:00 to 19:00 on one day, the output power curve and operating efficiency curve of the  battery system towards FFR can be acquired, which is shown as Fig.~\ref{m5case}~(a).


By comparing the two curves of  battery system output power and the real RegD signals during scheduling period, and calculating the mean relative error, it can be found that the battery system can track the curve of RegD signals with minor deviation.

It is obvious that the scheduling approach can maximize the utilization of each battery module during the scheduling horizon, so there is no obvious power shortage for the frequency response.

From the output power curves of each module, it can be observed that when the power demand for frequency response is relatively low, the modules filtered by the activation priority algorithm are activated first, while some modules remain inactive without any output. 
For example, modules 1 and 2, which have relatively low energy efficiency and high aging subgradients, are not activated.

Fig.~\ref{m5case}~(b) illustrates the process of SoC change and the SOC balance degree in each battery module within the system during scheduling.
The change trend shows that the scheduling method proposed in this work can effectively promote the SoC balance of each module while integrating them to join in FFR service with good performance.
The initial average error between the SoC of each battery module and the average SoC is 7.31\%, which is reduced to 4.93\% after 1-hour operation for FFR.

The aging subgradient curve of each module during scheduling is shown as Fig.~\ref{m5case}~(c).
It can be seen that the aging subgradient curve is closely related to each cycle during scheduling period and increases with the cycle depth. 
A comparison of this series of curves with the power curve in (a) shows that modules 1-4 have larger aging subgradients and, considering their lower energy efficiencies, participate less in responding to RegD signals.
Modules 5-8 have relatively lower aging subgradients and higher operation efficiencies, thus they contribute more power when the power demand for FFR is relatively high.
Modules 9 and 10 are relatively insensitive to cycle depth. 
At shallower cycle depth, their aging subgradients are higher compared to other lithium-ion battery modules, which can result in no power output. 
However, at deeper cycle depth, their aging sub-gradients are relatively low, leading to higher ouput power in response to FFR.

\subsubsection{Comparative Analysis}
In order to verify the effectiveness of the proposed scheduling approach, we compare it with two classical power allocation approaches applied for FFR: the maximum power-based power allocation method and the adjustable capacity-based power allocation method.
We also compare the scheduling results with those of the efficiency-aware method proposed in this work, where the aging subgradients are not weighted in the optimization objective.

For the maximum power-based power allocation method, it considers the maximum output power of each battery module and allocates the power for each module according to the ratio of its maximum output power to the system's maximum output power. 

For the adjustable capacity-based power allocation method, it considers the remaining adjustable capacity of each module during scheduling period, which depends on the charge/discharge command reflected by the RegD signals as well as the current SoC of the module. 

\begin{figure*}
    \centering
    \setlength{\abovecaptionskip}{0pt}
    \setlength{\belowcaptionskip}{0pt}
    \setlength{\floatsep}{0pt}
    \setlength{\textfloatsep}{0pt}
    \setlength{\intextsep}{0pt}
    \setlength{\parskip}{0pt}

    \subfigure[]{
        \begin{minipage}[t]{0.45\linewidth}
            \centering
            \includegraphics[width=3in]{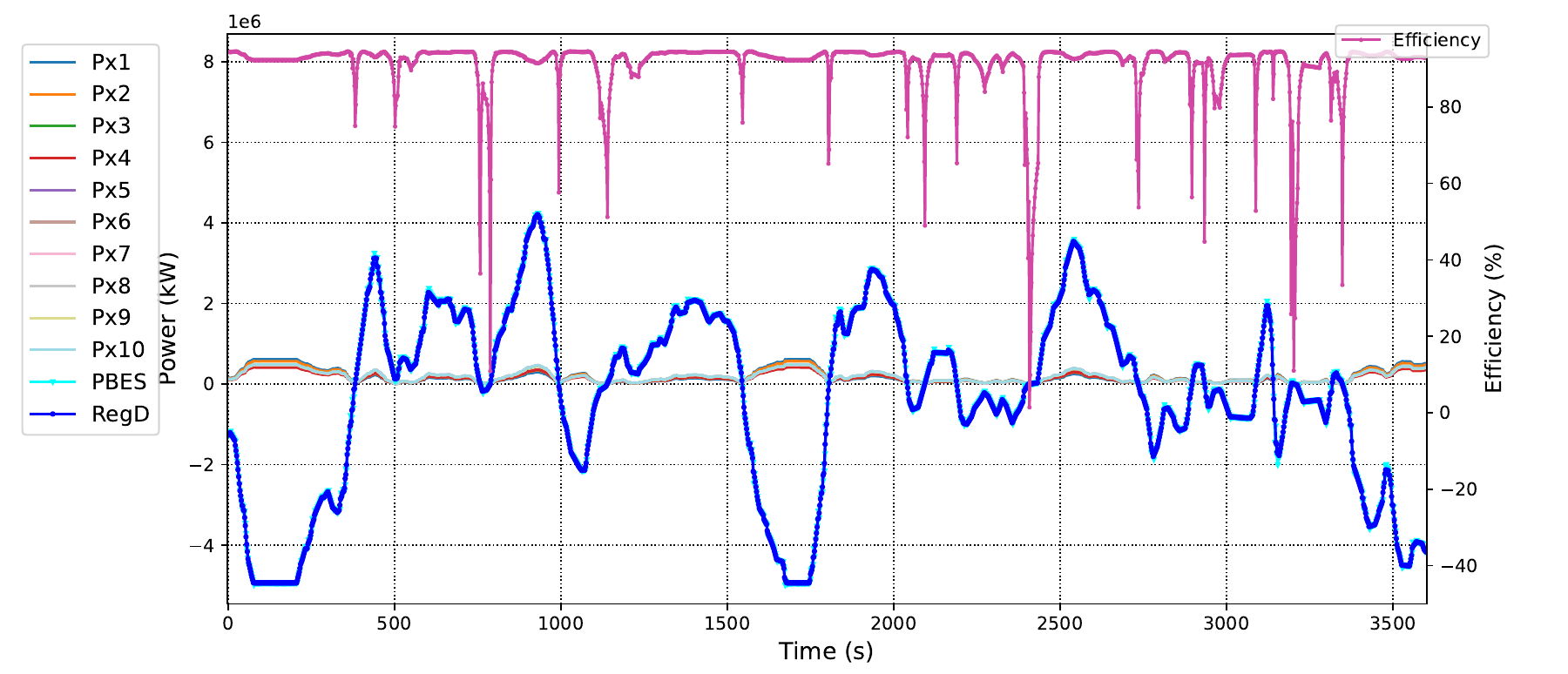}
            \vspace{-4.5mm} 
        \end{minipage}
    }%
    \hspace{2mm} 
    \subfigure[]{
        \begin{minipage}[t]{0.45\linewidth}
            \centering
            \includegraphics[width=3in]{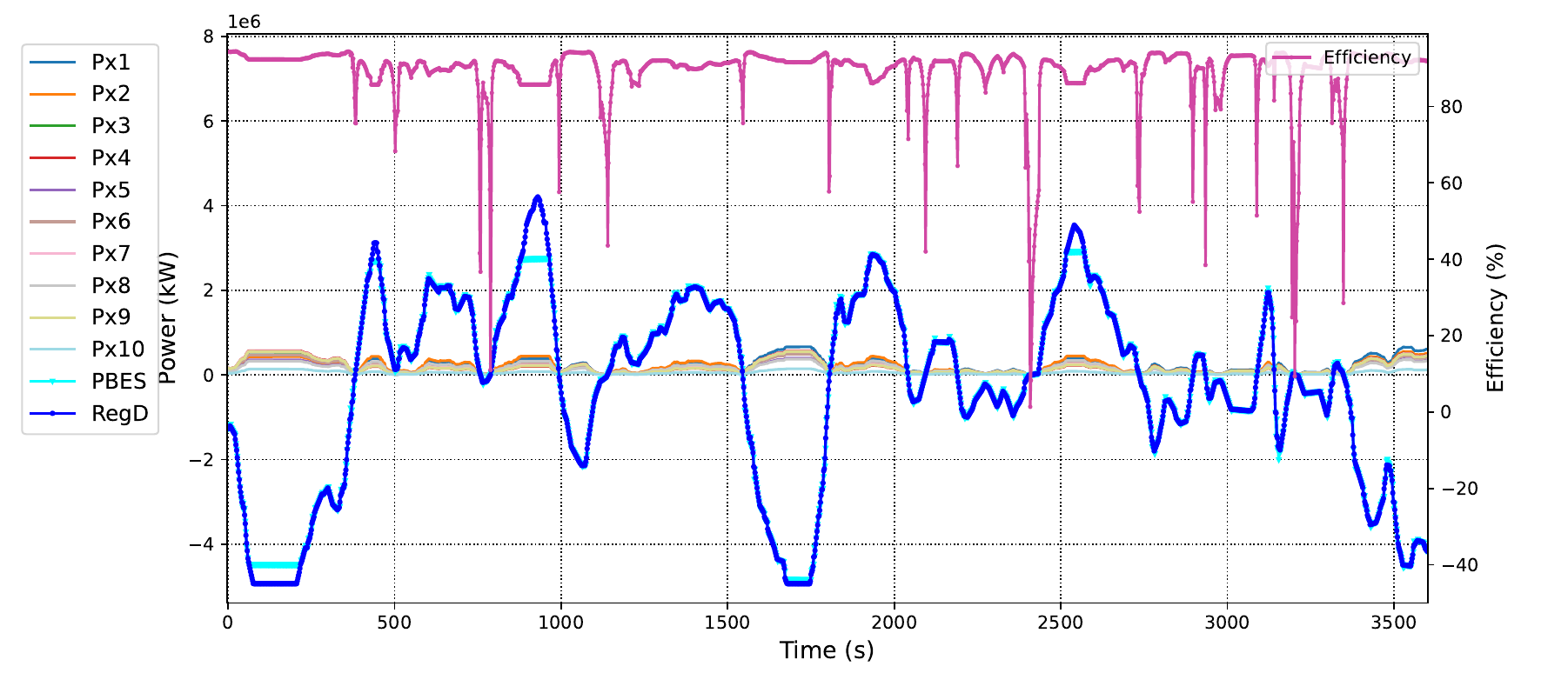}
            \vspace{-4.5mm}
        \end{minipage}
    }
    \vspace{-1mm} 
    \\
    \subfigure[]{
        \begin{minipage}[t]{0.45\linewidth}
            \centering
            \includegraphics[width=3in]{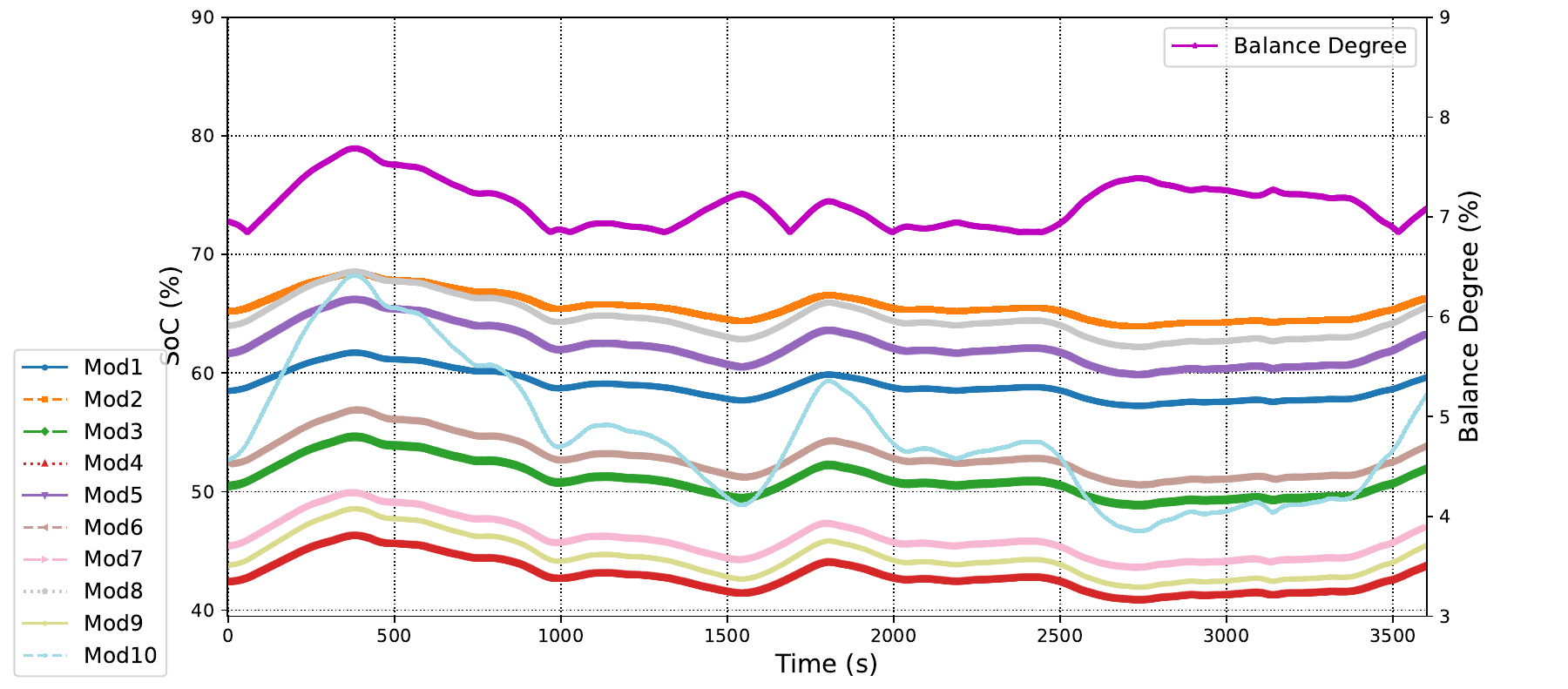}
            \vspace{-4.5mm}
        \end{minipage}
    }%
    \hspace{2mm}
    \subfigure[]{
        \begin{minipage}[t]{0.45\linewidth}
            \centering
            \includegraphics[width=3in]{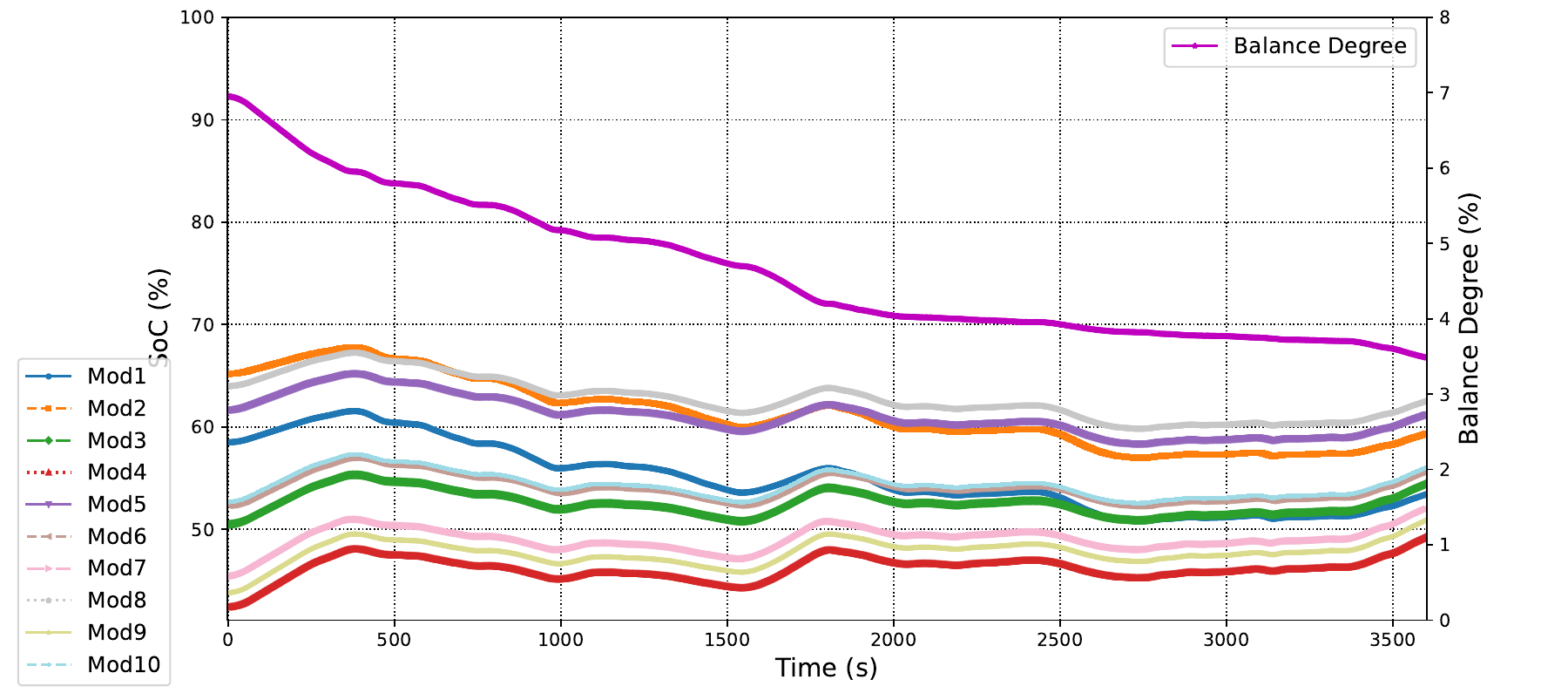}
            \vspace{-4.5mm}
        \end{minipage}
    }
    \vspace{-1mm}
    \\
    \subfigure[]{
        \begin{minipage}[t]{0.45\linewidth}
            \centering
            \includegraphics[width=3in]{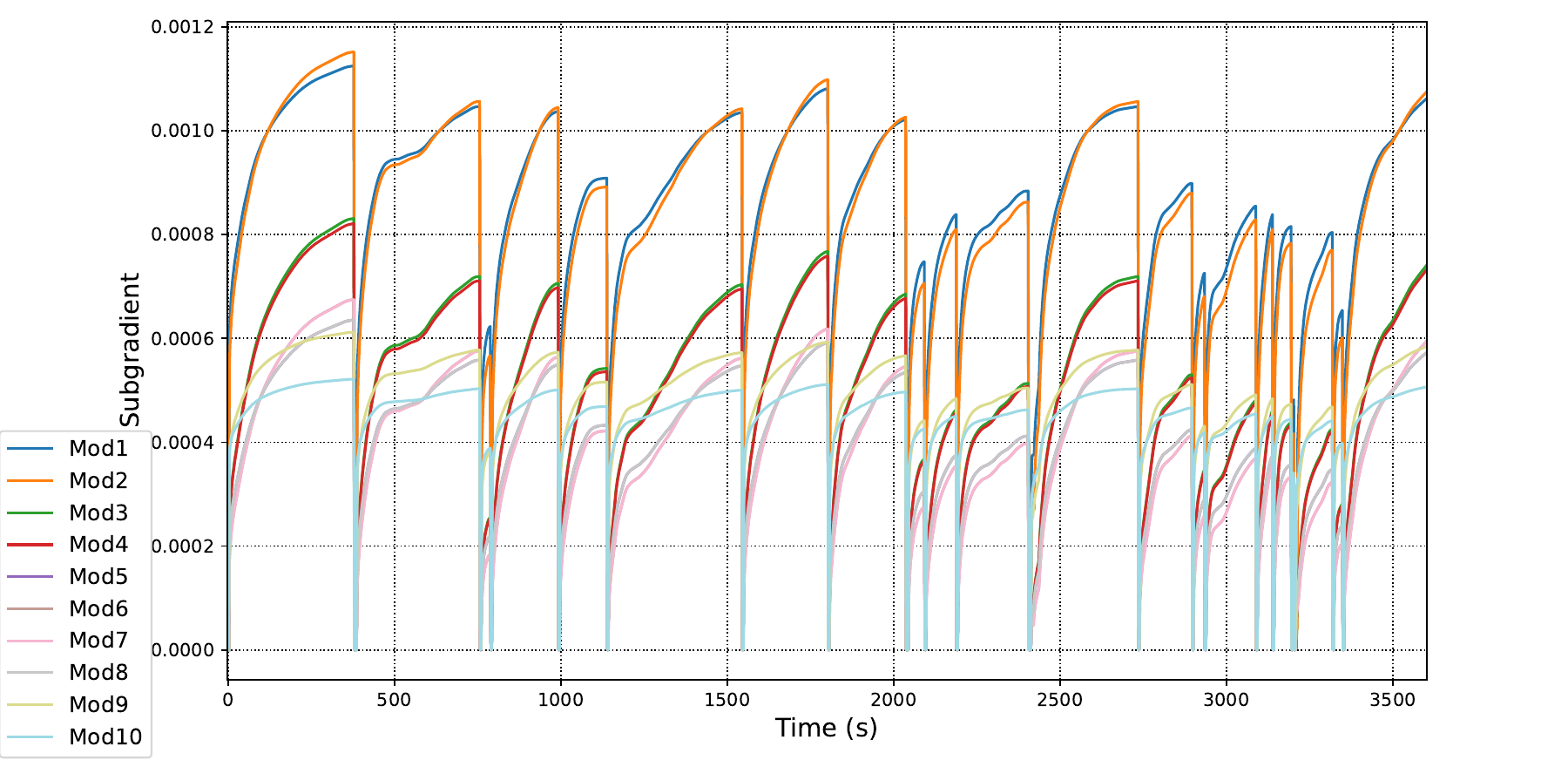}
            \vspace{-4.5mm}
        \end{minipage}
    }%
    \hspace{2mm}
    \subfigure[]{
        \begin{minipage}[t]{0.45\linewidth}
            \centering
            \includegraphics[width=3in]{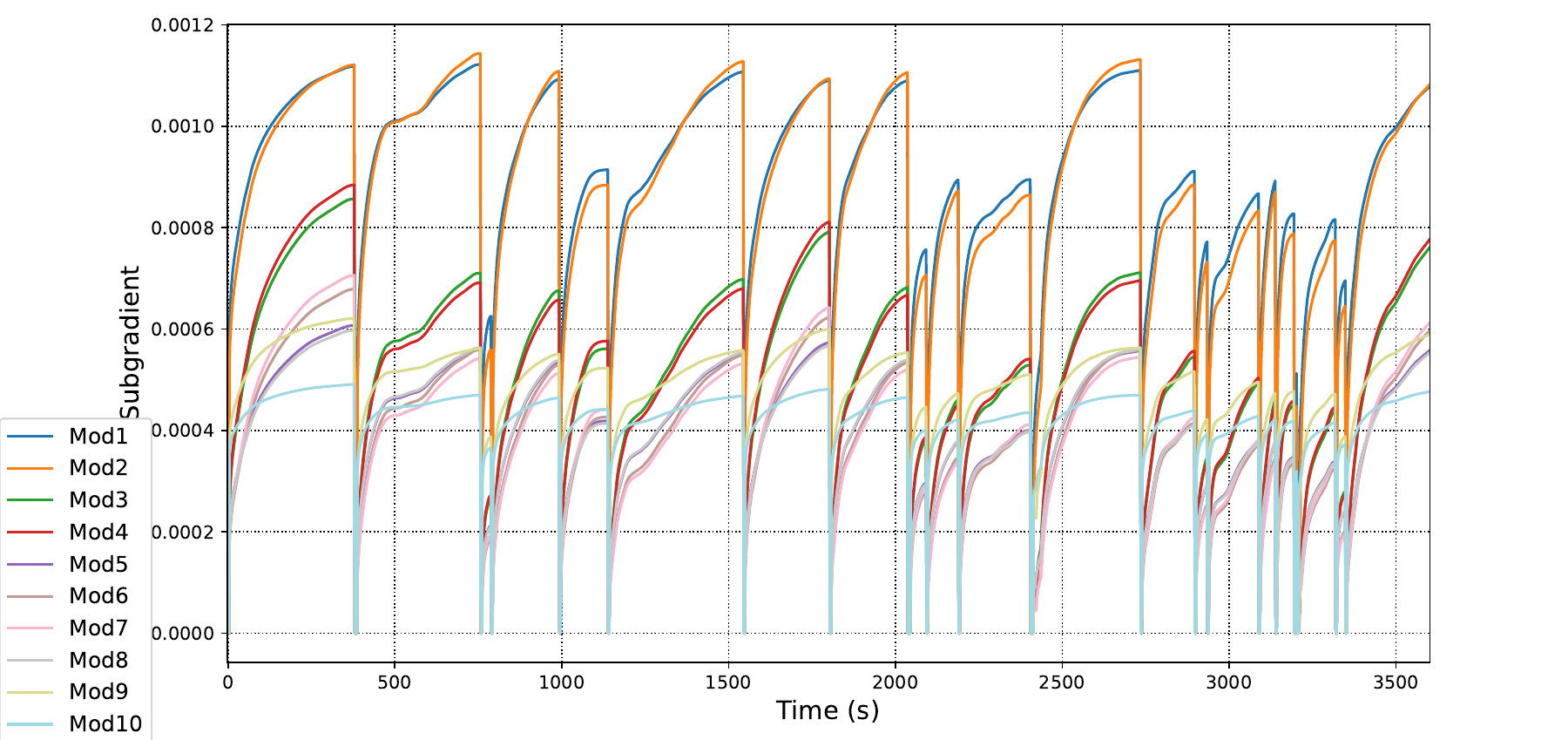}
            \vspace{-4.5mm}
        \end{minipage}
    }
    \caption{The scheduling results of the battery system using different methods in Case I.
        (a)~The power curve in Method 1. (b)~The power curve in Method 2.
        (c)~The SoC curve and balance degree in Method 1. (d)~The SoC curve and balance degree in Method 2.
        (e)~The aging subgradient curve in Method 1. (f)~The aging subgradient curve in Method 2.}
    \label{com_result25}
\end{figure*}

For ease of description, we refer to these two methods as Method 1 and Method 2 in this section.
These two methods are integrated in the MPC framework and the switching flexibility of each module is not taken into account, i.e., each module can participate in FFR at any scheduling interval. 
The scheduling results of the battery system towards FFR is illustrated by Fig.~\ref{com_result25}.
The comparison of scheduling results between the two comparative methods, the proposed performance-aware method, and the efficiency-aware method is presented by Table~\ref{tbl:comparison25}.

It can be observed that, with comparable prediction accuracy, both comparative methods show insufficient response during periods of high power demand for regulation, such as the intervals of 910-944s and 2510-2594s. 
In contrast, the proposed method reduces the regulation penalty cost by 33.16\% and 62.29\%, respectively, compared to the two methods, significantly improving the performance of the battery system for FFR service.

In terms of operation efficiency, the proposed method reduces energy loss cost by 51.69\% and 57.48\%, respectively, compared to the two methods, while ensuring good regulation performance. The overall operation efficiency is also comparable to the efficiency-aware method. 

With respect to battery aging when responding to FFR, the proposed method curtails aging cost by 14.10\% and 15.53\%, respectively, compared to the two methods, highlighting its significant role in alleviating the battery aging caused by frequent cycles of the modules during scheduling.
Compared to the efficiency-aware method, the proposed approach reduces aging cost by 3.31\% slightly. 
This is because, in this case, the lithium-ion battery modules with higher operation efficiencies also have longer lifespans, allowing the former to also contribute to optimizing aging cost.

To comprehensively evaluate the effectiveness of the proposed method over an extended timescale, a 12-hour simulation was conducted, spanning from 12:00 to 24:00 within a single day. A comparative analysis is also summarized in Table \ref{tbl:12h}.

It is evident that the proposed method also shows significant advantages for FFR over long horizon. In terms of energy efficiency, the proposed method reduces energy loss by 62.26\% and 51.22\% compared to Method 1 and Method 2, respectively. For the battery aging, the proposed method achieves a reduction of 17.59\% and 12.34\% in aging cost compared to Method 1 and Method 2, respectively.

\begin{table}
    \centering
    \caption{The scheduling results by different methods in case~I}
    \label{tbl:comparison25}
    \setlength{\tabcolsep}{0.28mm}{
    \begin{tabular}{c c c c c}
        \Xhline{2pt}
        \multirow{2}*{Results} & \multirow{2}*{\makecell[c]{Performance\\-aware}} & \multirow{2}*{\makecell[c]{Efficiency\\-aware}} & \multirow{2}*{\makecell[c]{Maximum\\power-based}} & \multirow{2}*{\makecell[c]{Adjustable\\capacity-based}} \\
        \\
        \Xcline{1-1}{0.4pt}
        \Xhline{1pt}
        Total energy throughput/kWh & 1763.04 & 1764.21 & 1748.71 & 1714.80 \\
        Total energy loss/kWh & 80.06 & 78.45 & 165.71 & 188.28 \\               
        Regulation penalty cost/\$ & 27.30 & 27.27 & 40.44 & 72.33 \\   
        Average prediction error/\% & 3.43 & 3.56 & 3.30 & 3.29 \\
        Aging cost/\$ & 920.11 & 942.23 & 1091.56 & 1092.58 \\
        Total running time/s & 3516.54 & 3403.24 & 3125.23 & 3158.45 \\  
        Average operation efficiency/\% & 95.66 & 95.74 & 90.32 & 90.43 \\ 
        SoC deviation reduction/\% & 2.38 & 2.37 & 0.84 & 2.32 \\ 
        \Xhline{2pt}
    \end{tabular}}
\end{table}



\begin{table}
    \centering
    \caption{The scheduling results by different methods of 12 hours in case~I}
    \label{tbl:12h}
    \setlength{\tabcolsep}{0.3mm}{
    \begin{tabular}{c c c  c}
        \Xhline{2pt}
        \multirow{2}*{Results} & \multirow{2}*{\makecell[c]{Performance\\-aware}}  & \multirow{2}*{\makecell[c]{Maximum\\power-based}} & \multirow{2}*{\makecell[c]{Adjustable\\capacity-based}} \\
        \\
        \Xcline{1-1}{0.4pt}
        \Xhline{1pt} 
        Total energy throughput/kWh & 15254.57 & 15817.82 & 15226.24 \\
        Total energy loss/kWh & 610.32 & 1616.7 & 1251.17 \\               
        Regulation penalty cost/\$ & 246.43  & 417.94 & 234.92 \\   
        Aging cost/\$ & 7852.97  & 9528.96  & 8958.40 \\
        Average operation efficiency/\% & 96.15  & 90.73 & 92.41 \\ 
        \Xhline{2pt}
    \end{tabular}}
\end{table}

\subsection{Case II:~The aggregation of multiple battery systems}

In this case, we consider to aggregate and schedule multiple BESSs with heterogeneous battery modules in regional grid to participate in the PJM regulation market.

\begin{figure}
\centering
\hspace*{0.5cm} 
\includegraphics[width=3.3in]{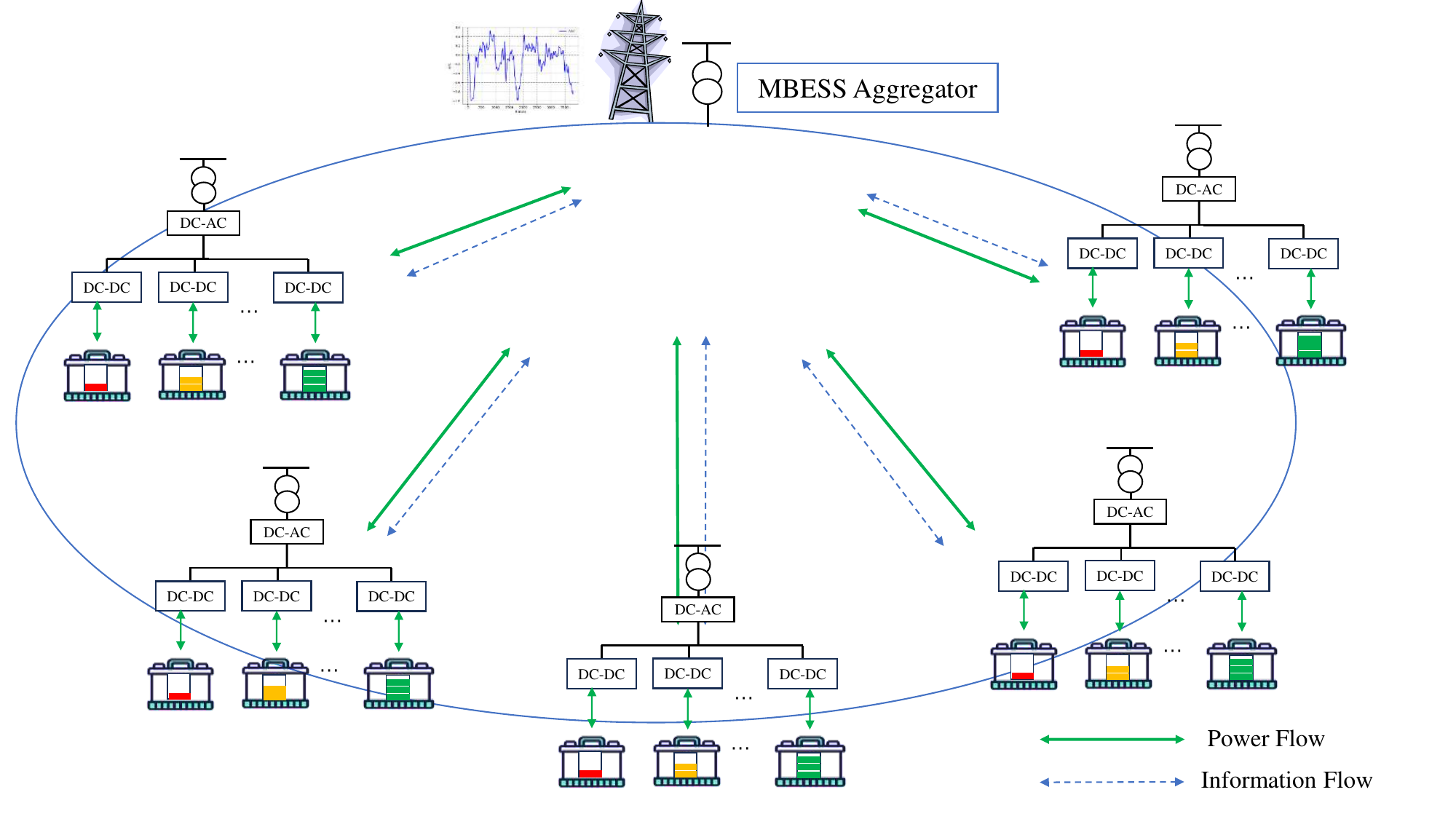}
\caption{The aggregation of multiple battery systems towards FFR}
\label{mbess}
\end{figure}

\subsubsection{Test System Description}

In this case, we set up that there are five modular lithium-ion battery energy storage systems in the regional distribution network, each of which consists of 30 modules connected in parallel on the DC side, as shown by Fig.~\ref{mbess}.
With respect to the operation range of the SoCs for each module, the settings for the FFR service in the PJM market and the settings for the MPC-based scheduling framework are similar to those in Case I.

\subsubsection{Simulation Results}

The offline optimization is firstly implemented according to the step size of ${\rm{s}} = 0.001$ and the initial condition of $SoC_{i,0}^{} = 0.5$ to obtain the lookup table for the MPC-based scheduling of multiple BESSs. 

\begin{figure}
	\centering
	\subfigure[]{
		\begin{minipage}[t]{0.5\linewidth}
			\centering
			\hspace*{-0.83in} 
			\includegraphics[width=3.1in]{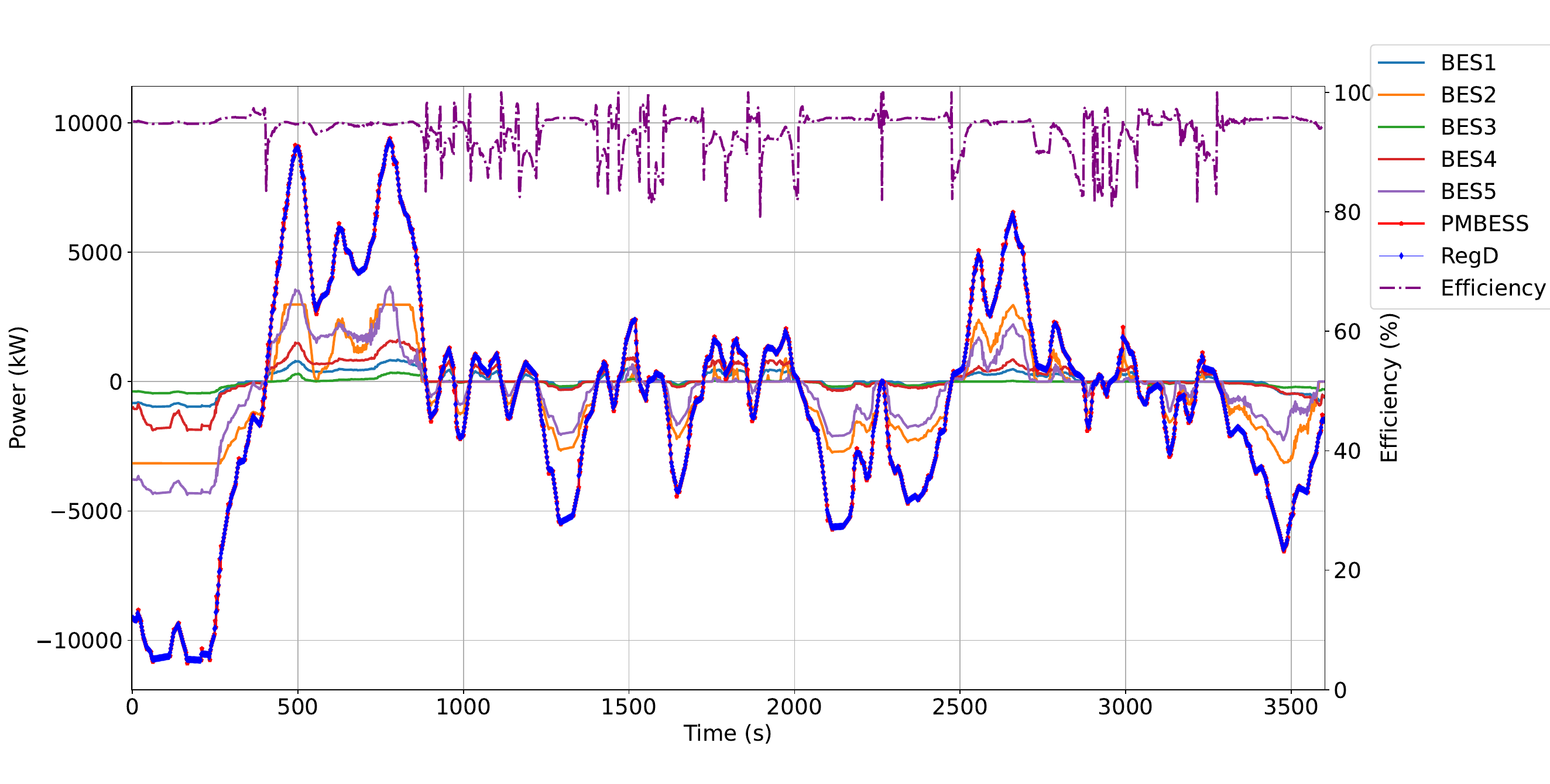}
		\end{minipage}
	}%
       \vspace{-3mm} 
       \\
	\subfigure[]{
		\begin{minipage}[t]{0.5\linewidth}
			\centering
			\hspace*{-0.9in} 
			\includegraphics[width=3.1in]{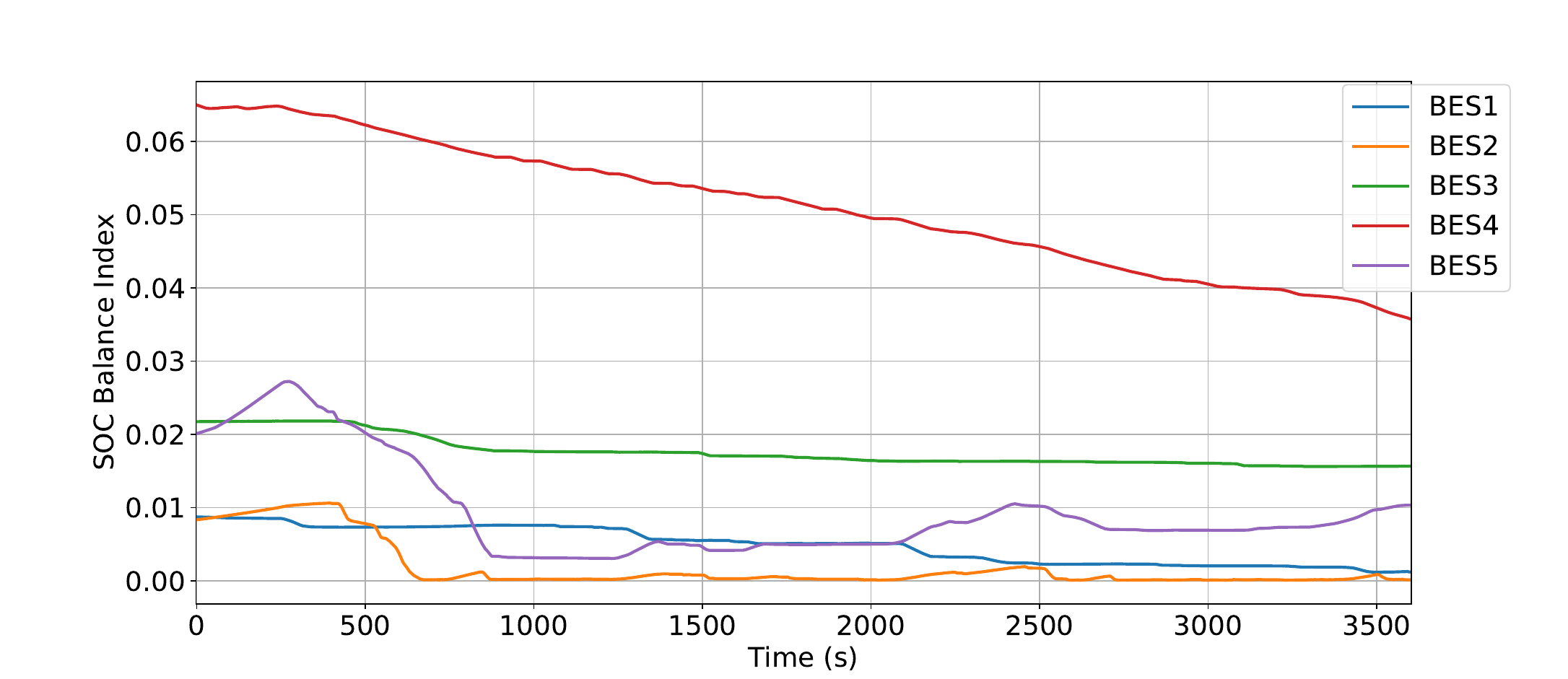}
		\end{minipage}
         }

	\centering
	\caption{The scheduling results of multiple battery systems in Case II.
 (a)~The power curve and the operation efficiency curve. (b)~The curve of SoC balance degree in Case II. }
	\label{150case_result}
\end{figure}

By simulating the participation of multiple battery systems in the PJM regulation market from 19:00 to 20:00 on one day, the power curve of each battery system and operation efficiency curve of each BESS participating in the FFR can be obtained, as shown in Fig.~\ref{150case_result}~(a).

It can be seen that the aggregator of five BESSs can satisfy the power demand of the FFR service well, and there is no obvious power shortage when the power demand is high. 
Only when the power demand of FFR is low, i.e., less than 0.5
\% of the rated power of the BESS aggregator, some battery modules are not activated for output because of the low operation efficiency at these operating points.

Fig.~\ref{150case_result}~(b) gives the variation curve of SoC balance degree in each BESS during the scheduling time period. 
Obviously, the SoC imbalance degree in each BESS is decreasing, which verifies that the Activation Priority Algorithm promotes the balance of modular SoC in each system while optimizing the energy loss cost and aging cost.

Similarly, the two power allocation methods, i.e., Method 1 and Method 2 mentioned in Case I are also utilized for the comparison in this case. 
The scheduling results using the proposed method are compared with those of the two comparative methods as shown in Table~\ref{tbl:comparison150}.
From the comparison of the scheduling results, it can be seen that the proposed approach reduces the total energy loss cost by 31.68\% compared to Method 1 and by 27.29\% compared to Method 2, while ensuring the responding quality of the five battery systems for FFR cooperatively.
Besides, the proposed method can cut down aging cost by 4.29\% compared to that of Method 1, while the aging cost of Method 2 is relatively low due to inadequate power response to RegD signals.
The result validates the scalability of the proposed approach for scheduling large-scale heterogeneous battery energy storage systems participating in FFR.




\begin{table}
    \centering
    \caption{The scheduling results by different methods in Case~II}
    \label{tbl:comparison150}
    \setlength{\tabcolsep}{0.45mm}{
    \begin{tabular}{c c c c}
        \Xhline{2pt}
         \multirow{2}*{Results} & \multirow{2}*{ \makecell[c]{Performance \\-aware }} & \multirow{2}*{ \makecell[c]{Maximum \\power-based}} & \multirow{2}*{ \makecell[c] {Adjustable\\capacity-based}}
         \\

        \\
   \Xcline{1-1}{0.4pt}
        \Xhline{1pt}

       Total energy throughput/kWh & 3096.98 & 3105.97 & 2491.08\\
       Total energy loss/kWh & 165.34 &  242.01  & 227.42  \\               
       Regulation penalty cost/\$& 7.52 &  74.29  & 654.79  \\   
        Average prediction error/\% & 4.14 &  4.13  & 4.13  \\
         Average operation efficiency/\%& 94.89 &  92.77 & 91.63  \\
        Aging cost/\$ &  1523.6 & 1592.02 & 1244.10 \\ 
     SoC deviation reduction/\%& 4.56 &  2.43  & 3.57   \\ 
        \Xhline{2pt}
    \end{tabular} }
\end{table}

\section{Conclusion}



This paper proposes a performance-aware MPC method for modular battery systems to enable FFR. The method optimizes economic profit in the regulation market while minimizing energy loss and battery aging cost. An electrical model quantifies dynamic losses, and a cycle-based aging model with real-time aging subgradient calculation evaluates aging cost during frequent cycling. The optimal scheduling problem is formulated as a MIQCP, solved using a real-time MPC framework with second-scale updates of RegD signals. To ensure computational efficiency, an activation priority algorithm combines offline optimization with online evaluation, accelerating the solution of the MIQCP problem.

Case studies show that the proposed scheduling approach can effectively reduce power loss cost by over 50\% in Case I and nearly 30\% in Case II, and cut down battery aging cost by nearly 15\% and 4.3\% respectively compared to conventional methods, which can also enhance the SoC balance. 
Future work will focus on extending this scheduling strategy to integrate frequency support with other grid services.
\bibliographystyle{ieeetr}
\bibliography{ref}

 
\vspace{22pt}

\begin{IEEEbiography}[{\includegraphics[width=1in,height=1.25in,clip,keepaspectratio]{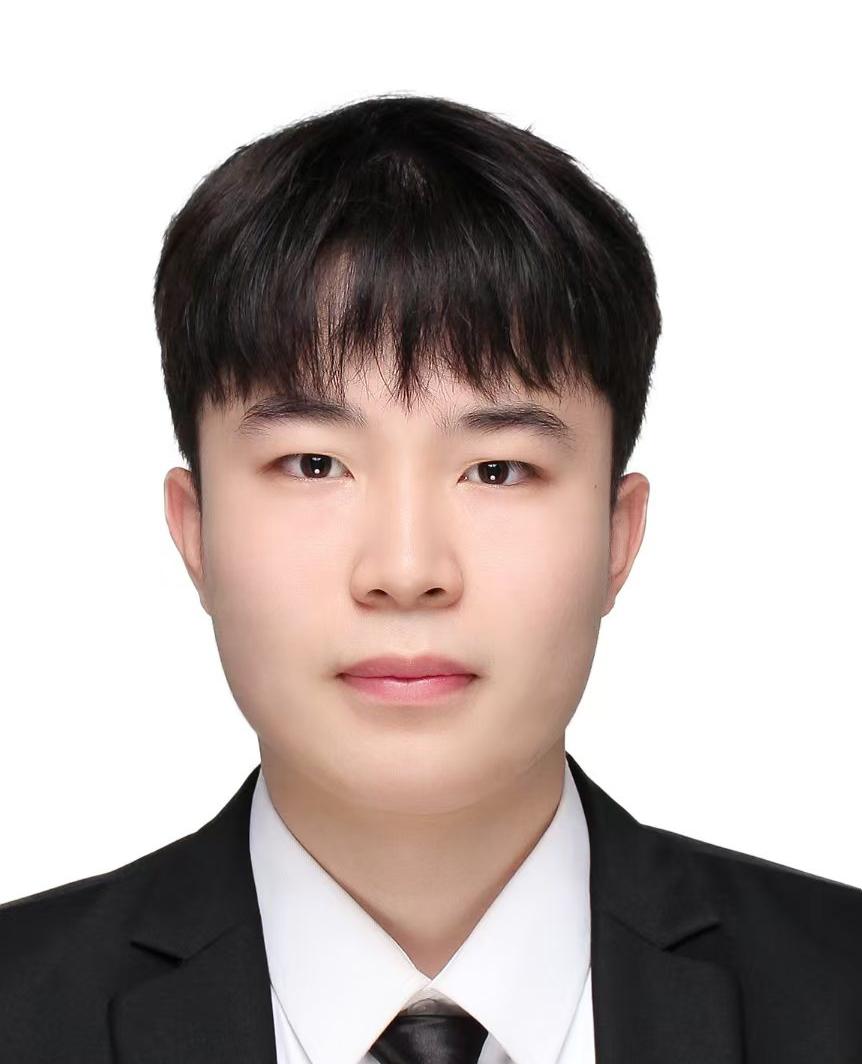}}]{Yutong He}
received the B.Sc. degree in electrical
engineering from North China Electric Power University, Baoding,
China, in 2022. He is currently working toward the
M.S. degree with the Department of Electrical Engineering, Tsinghua University, Beijing, China. His
research interests include modeling and operation of
battery energy storage systems and distribution system optimization.
\end{IEEEbiography}


\begin{center}
\begin{minipage}[t]{0.48\textwidth}
  \begin{IEEEbiography}[{\includegraphics[width=1in,height=1.25in,clip,keepaspectratio]{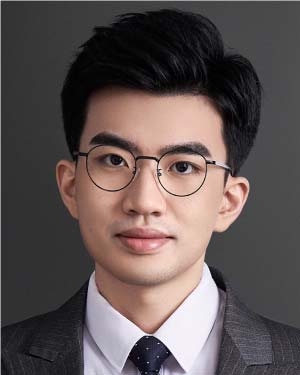}}]{Guangchun Ruan}
    (Member, IEEE) received the Ph.D. degree in electrical engineering from Tsinghua University, Beijing, China, in 2021. He visited the University of Washington, Seattle, WA, USA, in 2019, and Texas A\&M University, College Station, TX, USA, in 2020. He was a Postdoc with The University of Hong Kong, Hong Kong, in 2022. He is currently a Postdoc with the MIT Laboratory for Information and Decision Systems, Cambridge, MA, USA. His research interests include electricity market, energy resilience, demand response, data science, and machine learning applications.
  \end{IEEEbiography}
\end{minipage}
\hfill 
\begin{minipage}[t]{0.48\textwidth}
  \begin{IEEEbiography}[{\includegraphics[width=1in,height=1.25in,clip,keepaspectratio]{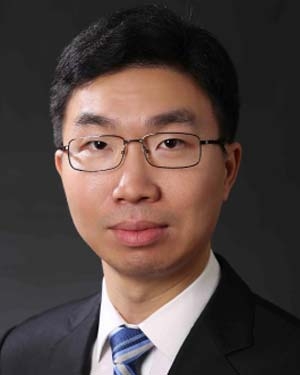}}]{Haiwang Zhong}
    (Senior Member, IEEE) received the B.S. and Ph.D. degrees in electrical engineering from Tsinghua University, Beijing, China. He is currently an Associate Professor with the Department of Electrical Engineering, Tsinghua University. He is also the Director of Energy Internet Trading and Operation Research Department, Sichuan Energy Internet Research Institute, Tsinghua University. His research interests include power system operations and optimization, and electricity markets.
  \end{IEEEbiography}
\end{minipage}
\end{center}

\end{document}